%% file: VVH-draft_final.tex
\documentclass [11pt,twoside,a4paper]{article}
\usepackage{shrthnds}
\usepackage{cite}
\usepackage{graphicx}
\usepackage{hyperref}
\hypersetup{
colorlinks,
linkcolor={blue},
citecolor={blue},
urlcolor={blue}} 
\usepackage{multirow}
\usepackage[usenames]{color}
\usepackage{subfigure}
\usepackage{setspace}
\usepackage[hang,small]{caption}
\usepackage{amsmath,amssymb}
\usepackage{slashed}   
\usepackage{float} 
\usepackage{placeins}
\restylefloat{table}             
\usepackage{geometry}
\usepackage{fancyhdr}
\newcommand{\auth}[2]{{\large #1}\footnote{email:~#2}}
\newcommand{\affS}[1]{{\it #1}}
\newcommand{\verttle}[2]{\vspace{-1cm}\begin{flushright}{\small #1}\end{flushright}\vspace{0.5cm} {\sf\bfseries #2}}
\newcommand{\nabstract}[2][]{\bc\begin{minipage}{0.9\textwidth}\begin{spacing}{1}{\small {\sf\bfseries Abstract:} #2 }\end{spacing}
#1 \end{minipage}\ec}
\newcommand{\nkeywords}[1][]{~\\\small{ {\sf\bfseries Keywords:} #1 }}

\usepackage{ulem}

\usepackage{color}
\definecolor{viv}{RGB}{255,109,254}

  \DeclareMathAlphabet{\pazocal}{OMS}{zplm}{m}{n}

\newcommand{\GGAAH}{gg \to \gamma \gamma H}
\newcommand{\GGAZH}{gg \to \gamma ZH}
\newcommand{\GGZZH}{gg \to ZZH}
\newcommand{\GGWWH}{gg \to WWH}
\newcommand{\tin}[1] {\textit{\tiny #1}}

\title{\verttle{}{Di-vector boson production in association with a Higgs boson at hadron colliders \\}}
\author{
  \auth{Pankaj Agrawal$^{a,b,c}$}{agrawal@iopb.res.in}~,
  \auth{Debashis Saha$^{a,b}$}{debasaha@iopb.res.in}~, and
  \auth{Ambresh Shivaji$^d$}{ashivaji@iisermohali.ac.in}\\~\\  
  \affS{a) Institute of Physics, Sainik School Post, Bhubaneswar 751005, India}\\
  \affS{b) Homi Bhabha National Institute, Training School Complex, }\\
   \affS{ Anushakti Nagar, Mumbai 400085, India}\\
  \affS{c) Department of Physics, Indian Institute of Technology Delhi,}\\ 
     \affS{ Hauz Khas, New Delhi-110016, India }\\  
  \affS{d) Indian Institute of Science Education and Research, Knowledge City,} \\
  \affS{Sector 81, S. A. S. Nagar, Manauli PO 140306, Punjab, India}\\~\\
}
\newcommand{\pghdr}{}
\date{\today}

\begin{document}
\vspace{1.5in}
\maketitle

\nabstract[\nkeywords{Electroweak, Higgs boson, Polarization, LHC, Anomalous couplings}]
{
We consider the production of a Higgs boson in association with two electroweak vector bosons at hadron colliders. In particular, 
we examine $\gamma\gamma H$, $\gamma ZH$, $ZZH$, and $W^{+}W^{-}H$ production at the LHC (14 TeV), HE-LHC (27 TeV), and 
FCC-hh (100 TeV) colliders. Our main focus is to estimate the gluon-gluon ($gg$)  channel ($gg \to VV^\prime H$) contributions to 
$pp \to VV^\prime H~(V,V^\prime=\gamma,Z,W)$ and compare them with corresponding contributions arising from the quark-quark ($q q$)  channel ($q {\bar q} \to VV^\prime H$). Technically, 
the leading order $gg$ channel contribution to $pp \to VV^\prime H$ cross section is a next-to-next-to-leading order  
correction in strong coupling parameter, $\alpha_s$.
In the processes under consideration, we find that in the $gg$  channel, $W^{+}W^{-}H$ has the largest cross section. However, relative contribution of the $gg$ channel is more important for the $pp \to ZZH$ production. At the FCC-hh, $gg \to ZZH$ contribution is comparable with
the next-to-leading order QCD correction to $qq \to ZZH$. 
We also compute the cross sections when $W$ and $Z$-bosons are polarized. In the production of $W^{+}W^{-}H$ and $ZZH$, we find that the $gg$ channel contributes more significantly when the vector bosons are longitudinally polarized. By examining such events, one
can increase the fraction  of the $gg$ channel contribution to these processes.
Further, we have studied beyond-the-standard-model effects in these processes using the $\kappa$-framework parameters 
$\kappa_t, \kappa_V$, and $\kappa_\lambda$. We find that the $gg$  channel processes $ZZH$ and $WWH$ have very mild 
dependence on $\kappa_\lambda$, but strong dependence on $\kappa_t$ and $\kappa_V$. The $qq$ channel processes mainly depend on $\kappa_V$. Dependence of the $gg$ channel contribution on $\kappa_V$ is stronger than that of the $qq$ channel contribution.
Therefore focusing on events with longitudinally polarized $W$ and $Z$-bosons, one can find stronger dependence on  $\kappa_V$ that can help us measure this parameter. 
} 

\newpage

 \section{Introduction}\label{sec:intro}

    After the discovery of a Higgs-like resonance, with a mass of 125 GeV, at the Large Hadron Collider (LHC) in 2012, various properties of this new particle have been studied. The spin and parity measurements have established it as a $0^{+}$ state at 99.9\% CL against  alternative scenarios~\cite{Aad:2015mxa}. Couplings of this new particle with the fermions and gauge bosons predicted in the standard model are getting constrained as more and more data 
are being analyzed by the LHC experiments~\cite{Khachatryan:2016vau,Sirunyan:2018koj,ATLAS:2019slw}. To this end, the vector-boson fusion production of the Higgs boson, associated production of $VH (V=Z,W)$, and Higgs boson's decay into vector bosons set limits on the $HVV$ couplings~\cite{Anderson:2013afp,CMS:2020dkv}. The gluon-gluon ($gg$) channel production of the Higgs boson helps in constraining the $H t\bar{t}$ coupling~\cite{CMS:2020dkv}. In addition, the evidence for the associated production of Higgs boson with a top-quark pair~\cite{Aaboud:2017jvq,Sirunyan:2018hoz} will provide the direct measurement of $Ht\bar{t}$ coupling. We still need to measure the trilinear and quartic Higgs self-couplings in order to know the form of the Higgs potential which will in turn reveal the exact symmetry breaking mechanism. 
The Higgs self-couplings can be probed directly in multi-Higgs production processes~\cite{TheATLAScollaboration:2014scd,Chen:2015gva,Fuks:2017zkg}. Recently, indirect methods of probing them at hadron 
and lepton colliders have also been proposed~\cite{Rossi:2020xzx,McCullough:2013rea,Borowka:2018pxx}.
Similarly, the quartic couplings involving Higgs and vector bosons $HHVV$ are also not constrained independently. This coupling can be probed in the vector-boson fusion production of a Higgs boson pair~\cite{Bishara:2016kjn,Aad:2020kub}. In order to find the signals of new physics, it is important that we improve our theoretical predictions for the processes 
involving Higgs boson at current and future colliders.
\\

Loop-induced decay and scattering processes can play an important role in searching for new physics. In the presence of new physics (new particles and/or interactions), the rates for such processes can differ significantly from their standard model predictions. In this regard, many $gg$ channel scattering processes in $2 \to 2$ and $2 \to 3$ category have been studied~\cite{deFlorian:1999tp, Melia:2012zg, Agrawal:2012df, Campanario:2011cs, Agrawal:2012as,
Campanario:2012bh, Shivaji:2013cca, Campbell:2014gua, Agrawal:2014tqa, Mao:2009jp, Hespel:2015zea, Gabrielli:2016mdd, Caola:2015psa, Caola:2015rqy, Campbell:2016yrh, Caola:2016trd, Granata:2017iod, Shivaji:2016lnu, Plehn:2005nk, Binoth:2006ym, Fuks:2017zkg, Maltoni:2014eza, Papaefstathiou:2015paa, Kilian:2017nio, Hirschi:2015iia,Agrawal:2017cbs}. In the present work, we are interested in loop-induced $gg$ channel contribution to $VVH$ ($\gamma \gamma H,~ \gamma ZH,~ ZZH$, and $W^{+}W^{-}H$) production. In QCD perturbation 
  theory, the leading order $gg$ channel contribution to $pp \to VVH$ is an NNLO contribution at the cross section level. Due to many 
  electroweak couplings involved and loop-induced nature of $gg \to VVH$ processes, their cross sections are expected to be small. 
  However, they can be important at high energy hadron colliders like 100 TeV $pp$ collider such as proposed hadronic 
  Future Circular Collider (FCC-hh) facility at 
  CERN~\cite{Mangano:2270978} and Super Proton-Proton Collider (SPPC) facility in China~\cite{CEPC-SPPCStudyGroup:2015csa}. 
  At such energy scale, the gluon flux inside the proton becomes very large.   
  In fact, for $\gamma\gamma H$, the $gg$ channel gives the dominant contribution. 
\\

Unlike the quark-quark contributions, which are mainly sensitive to $HVV$ 
couplings, the gluon-gluon contribution allows access to $Ht{\bar t}, HHH$, and $HHVV$ couplings as well.
Note that the processes under consideration are background to $pp \to HH$ when one of the Higgs bosons decays into $\gamma\gamma/ \gamma Z/ ZZ^*$
  or $WW^*$ final states. The process $pp \to ZZH$ is also a background to $pp \to HHH$ when two of the three Higgs bosons decay into
  $b{\bar b}$ final states.   
In this work, we present a detailed study of $gg \to \gamma\gamma H$ and $\gamma ZH$ for  the first time in the SM.   
The $gg$ channel contribution to $ZZH$ and $WWH$ in the SM have been studied in the past~\cite{Mao:2009jp,Baglio:2015eon,Baglio:2016ofi}. 
We have presented the $ZZH$ and $WWH$ calculations in detail and have proposed
methods to enhance the relative contribution of gluon-gluon 
channel over quark-quark channel. 
Since loop-induced processes are sensitive to new physics, 
we also study the effect of new physics in all $VVH$ processes using a common BSM framework --- the $\kappa$-framework. Going beyond the $\kappa$-framework, we have treated the $HHVV$ coupling independently and emphasized 
its effect in $ZZH$ and $WWH$ processes. BSM study in a more sophisticated 
framework is desirable but it is beyond the scope of the present work.
\\

 Experimentally, $W$ and $Z$-boson polarizations have been measured at hadronic
 colliders \cite{Chatrchyan:2011ig,Aad:2012ky,Aaboud:2019gxl}. We also compute the cross sections for the processes when
 these bosons are polarized. For each process, the different production channels contribute predominantly to specific polarization configurations. This can help in enhancing
 the contribution of the $gg$ channel, as compared to the $qq$ channel. The
 $gg$ channel have sometimes stronger dependence on the kappa parameters,
 in particular on $\kappa_V$. Therefore, an event sample with larger $gg$ channel
 contribution can be helpful.
\\

The paper is organized as follows.  In Sec.~\ref{sec:gg-fuse-VVH}, we discuss the Feynman diagrams which contribute to $gg \to VVH$ 
amplitudes. The model independent framework to study new physics effects is outlined in Sec.~\ref{sec:bsm-para}. In Sec.~\ref{sec:calc-check}, we provide details on the calculation techniques and various checks that we have performed in order to ensure the correctness of our calculation. In Sec.~\ref{sec:numrRes}, we present numerical results in SM and BSM scenarios for all the $VVH$ processes. Finally, we summarize our results and conclude in Sec.~\ref{sec:concl}.

\section{Gluon fusion Contribution to $VVH$}
\label{sec:gg-fuse-VVH}

\begin{figure}[h]
\includegraphics[angle=0,width=1\linewidth]{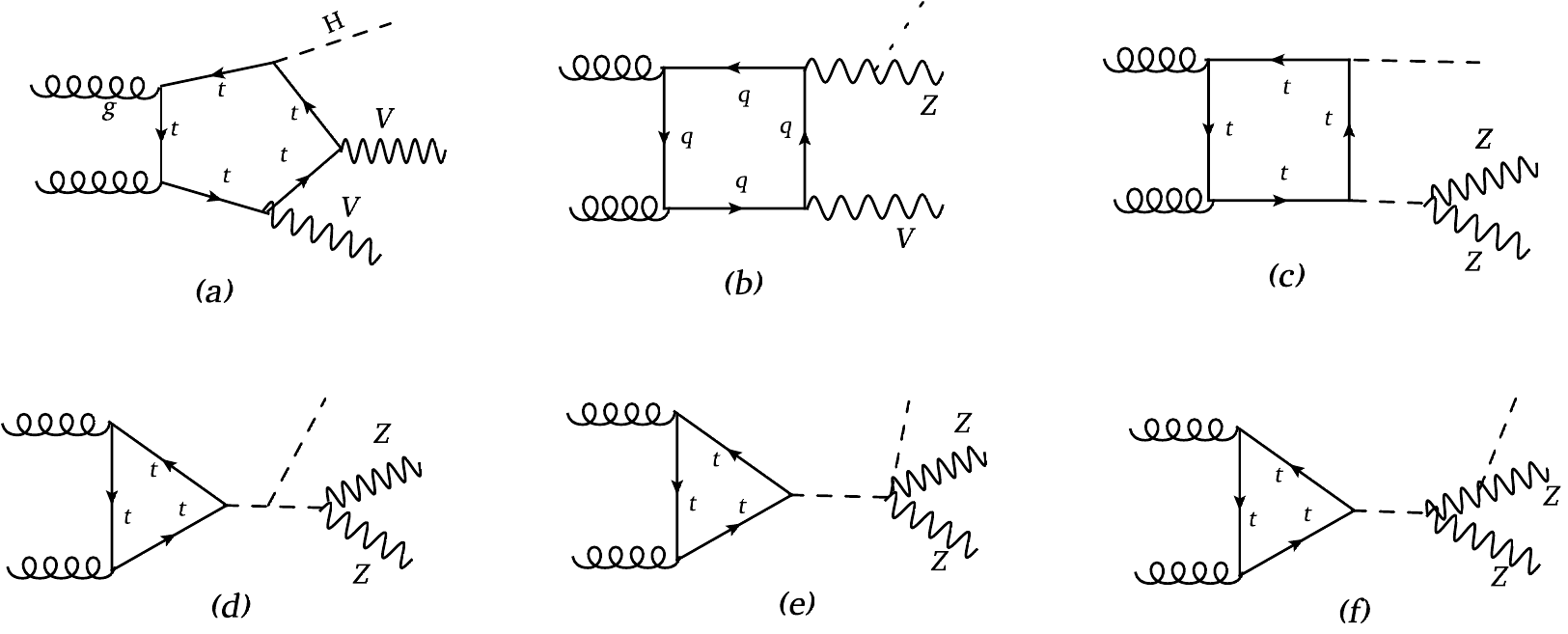}\\
	\caption{Different classes of diagrams for $gg \to VVH,~{V=\gamma, Z}$. In diagram (b), $q$ represents all quark flavors. 
	Process $gg \to \gamma\gamma H$ receives contribution only from (a) type diagrams, while $gg \to \gamma ZH$ gets contribution from both (a) and (b) type diagrams. In the case of $ZZH$, all the diagrams contribute; the diagrams (b) and (f) cover the situation in which $H$ is attached to a $Z$ boson. }
	\label{fig:feyn-A-Z-H-pen-bx-tr}
\end{figure}

\begin{figure}[H]
\includegraphics[angle=0,width=1\linewidth]{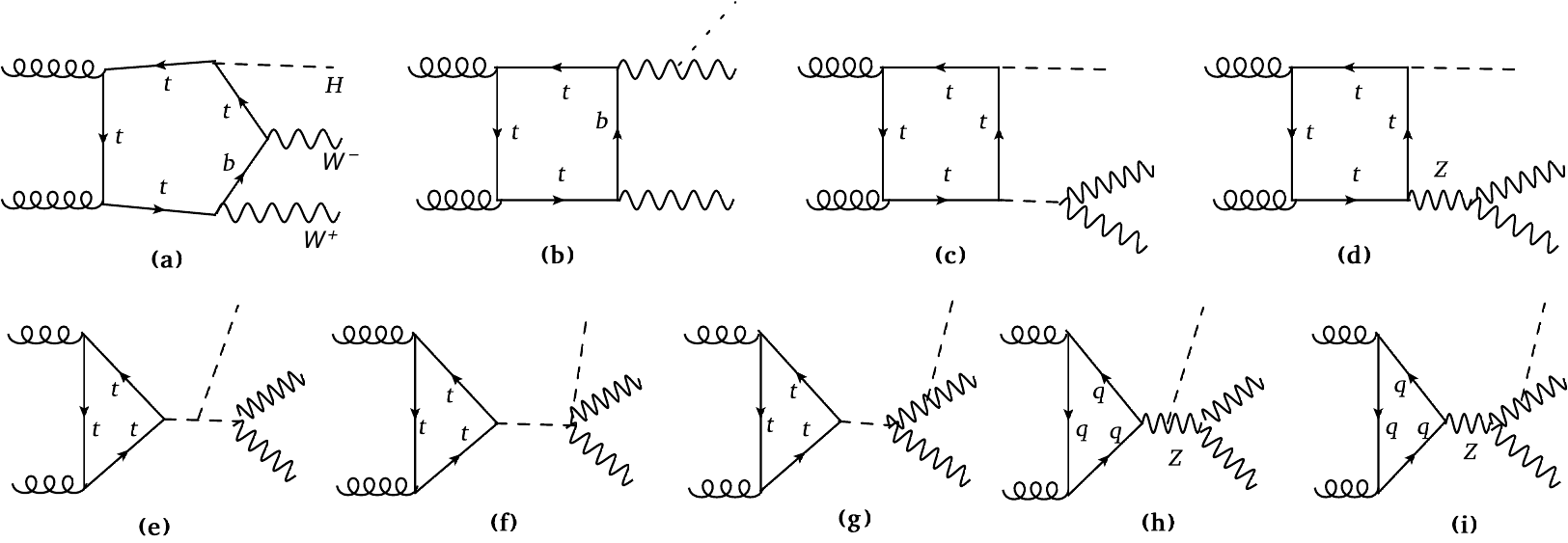}\\
	\caption{Different classes of diagrams contributing to $gg \to WWH$ process. With respect to $ZZH$, new 
	classes of box and triangle diagrams appear due to $ZWW$ 
	coupling. In (a) and (b), due to the flavor changing interaction of $W$ with quarks, both the quark 
	flavors of a given generation enter in the loop. The diagrams (b), (g) and (i) cover the case when 
	$H$ is attached to a $W$ boson. }
\label{fig:feyn-WWH-pen-bx-tr}
\end{figure}

The $gg$ channel contribution to $pp \to VVH$ is due to a loop-induced scattering process mediated by a quark-loop. The classes of diagrams contributing to $gg \to VVH$ processes are shown in Fig.~\ref{fig:feyn-A-Z-H-pen-bx-tr}\footnote{Feynman diagrams have been made using Jaxodraw~\cite{Binosi:2008ig}.}. 
For convenience, the diagrams contributing to $gg \to WWH$ process are shown separately in Fig.~\ref{fig:feyn-WWH-pen-bx-tr}. The $gg \to \gamma\gamma H$ process receives contribution only from the pentagon diagrams, while, $\gamma ZH$ receives contribution from both 
pentagon and box class of diagrams. In case of $gg \to ZZH,~WWH$ processes, triangle class of diagrams also contribute.  
We have taken all quarks but the top-quark as massless. Therefore, the top-quark contribution is relevant in 
diagrams where Higgs boson is directly attached to the quark loop. 
In the diagrams where Higgs boson does not directly couple to the quark loop, light quarks can also contribute.  
The complete set of diagrams for each process can be obtained by permuting external legs.
These permutations imply that there are 24 diagrams in pentagon topology, 6 diagrams in 
each box topology and 2 diagrams in each triangle topology.
The diagrams in which only one type of quark flavor runs
in the loop, are not independent. 
Due to Furry's theorem only half of them 
  are independent~\cite{10.1143/ptp/6.4.614}. This observation leads to a significant simplification in the overall calculation. 
This simplification, however, is not applicable to the $WWH$ case, where flavor changing interaction is 
involved in the quark loop. For example, see (a) and (b) in Fig.~\ref{fig:feyn-WWH-pen-bx-tr}.
\\

  Thus, there are 12 independent pentagon diagrams (Fig.~\ref{fig:feyn-A-Z-H-pen-bx-tr}(a)) due to top-quark loop 
  contributing to $gg \to \gamma\gamma H$ process. 
  Similarly, the $gg \to\gamma ZH$ process receives contribution from 12 independent pentagon diagrams (Fig.~\ref{fig:feyn-A-Z-H-pen-bx-tr}(a)) due to top-quark loop and 3 independent box 
  diagrams (Fig.~\ref{fig:feyn-A-Z-H-pen-bx-tr}(b)) for each quark flavor. In principle, 5 light quarks ($u, d, c, s, b$) and 1 heavy quark ($t$) 
  contribute. The box class of diagrams arise due to $ZZH$ coupling and has effective box topology of $gg \to \gamma Z^*$ 
  amplitude. Furry's theorem, in this case, implies that the axial vector coupling of $Z$ boson with quark does not 
  contribute to $gg \to \gamma ZH$ amplitude. 
\\    

  Like the $gg \to \gamma ZH$ process, the $gg \to ZZH$ amplitude receives contribution from 12 independent pentagon diagrams with top-quark in the loop
  (Fig.~\ref{fig:feyn-A-Z-H-pen-bx-tr}(a)). However, there are 6 independent box diagrams with effective box topology of $gg \to ZZ^*$ amplitude for each quark flavor 
  which covers the possibilities of $H$ coupling with any of the two external $Z$ bosons (Fig.~\ref{fig:feyn-A-Z-H-pen-bx-tr}(b)). 
  Further, a new box type contribution arises which has effective box topology of $gg \to HH^*$ amplitude 
  (Fig.~\ref{fig:feyn-A-Z-H-pen-bx-tr}(c)). Once again there are 3 such independent diagrams with only top-quark in the loop. In addition to that, 
  there are 4 independent triangle diagrams with top-quark in the loop and having effective triangle topology 
  of $gg \to H^*$ amplitude (Fig.~\ref{fig:feyn-A-Z-H-pen-bx-tr} (d), (e), (f)). In $gg \to ZZH$ amplitude, the Furry's theorem implies that the vector and axial vector coupling of $Z$ boson with quarks can contribute at  quadratic level only. 
\\

  Among all $VVH$ amplitudes, the structure of $gg \to WWH$ amplitude is the most complex. Due to the involvement 
  of flavor changing interactions in Fig.~\ref{fig:feyn-WWH-pen-bx-tr} (a) and (b), the Furry's theorem is not applicable to these diagrams.
  Therefore, 24 independent pentagon diagrams contribute to $gg \to WWH$ process for each generation of quarks. However, 
  since we neglect Higgs coupling with light quarks including the $b$ quark, there are only 12 non-zero independent 
  pentagon diagrams. In Fig.~\ref{fig:feyn-WWH-pen-bx-tr} (b), all the three quark generations contribute. Taking into account the possibility 
  of Higgs boson coupling with any of the two external $W$ bosons, there are total 12 independent box diagrams of type (b) for each generation.
  {In diagrams (a) and (b), the axial vector coupling of $W$ with quarks contributes at quadratic as well as at linear level.}
  Like in the $gg \to ZZH$ process, there are 3 independent box diagrams of type (c). Due to $ZWW$ coupling, a new box contribution of type (d) having 
  effective box topology of $gg \to HZ^*$ amplitude appears. Furry's theorem for diagram (d) implies that the vector coupling 
  of $Z$ with quarks does not contribute to the amplitude. The same explains the absence of similar box diagram due to 
  $\gamma WW$ coupling. Further, there are 4 independent triangle diagrams with top-quark loop (Fig.~\ref{fig:feyn-WWH-pen-bx-tr} (e), (f) (g)) as in case 
  of the $gg \to ZZH$ process. A new type of 3 independent triangle diagrams for each quark flavor with effective triangle topology of 
  $gg \to Z^*$ amplitude appears, once again due to $ZWW$ coupling 
  (Fig.~\ref{fig:feyn-WWH-pen-bx-tr} (h), (i)). These triangle diagrams are anomalous and they can receive contribution only from the third generation 
  quarks as the bottom and top-quarks have very different masses. This is indeed the case for (h) type diagrams. 
  However, we find that (i) type diagrams do not contribute. This is explained in the appendix A.

\section{BSM Parametrization}
\label{sec:bsm-para}

Measuring the couplings of the Higgs boson with fermions, gauge bosons and with itself is an important 
aspect of finding the signatures of new physics at colliders. With the help of the data collected 
so far at the LHC, we now know couplings of the Higgs boson with the top quark with an accuracy of 10-20\% and with vector bosons with an accuracy of 10\% at 1$\sigma$ ~\cite{Tanabashi:2018oca}. The Higgs self-couplings, on the other hand, 
are practically unconstrained~\cite{Aad:2019uzh}.
\\

To study the new physics effects in $VVH$ processes, we take the simplest approach of modifying the SM 
like couplings only, also known as the kappa framework for the parametrization of new physics~\cite{LHCHiggsCrossSectionWorkingGroup:2012nn,Ghezzi:2015vva}. 
In this framework, no new Lorentz structures and no new interaction vertices appear. The LHC 
experiments have interpreted the data using this framework so far.
The couplings of our interest are $Ht{\bar t}$, $HVV$, $HHH$ and $HHVV$. Out of these couplings, $gg \to \gamma\gamma H$ 
is sensitive to only $Ht{\bar t}$ coupling. The $HVV$ coupling affects all other processes. The couplings $HHH$ and  $HHVV$ affect only $gg \to VVH,~ {V=Z,W}$ processes.
\\  

The modification in these couplings due 
to new physics is implemented through scale factor $\kappa_{i }$ for various couplings of the Higgs boson in the SM. 
In kappa framework, there 
are three such scale factors namely $\kappa_t$ for Higgs coupling with top-quark, $\kappa_V$ for Higgs coupling with 
vector bosons ($\kappa_{HZZ} = \kappa_{HWW} = \kappa_V$)~\footnote{Note that in the SM, the tree level interaction vertices 
$H\gamma\gamma$ and $H\gamma Z$ do not exist.} and $\kappa_\lambda$ for Higgs coupling with itself.  
Since in the SM both $HVV$ and $HHVV$ couplings are related, the scaling of $HHVV$ coupling is also parametrized by $\kappa_V$. In a more generic BSM framework, the $HHVV$ coupling, in principle, can be independent of $HVV$ coupling.
\\

In the presence of BSM effects, the amplitudes for the $gg$ channel processes depend on $\kappa_t$, $\kappa_V$, and $\kappa_\lambda$
as follows.
\begin{eqnarray}
	{\cal M}^{\rm BSM}(gg \to \gamma\gamma H) &=& \kappa_t {\cal M}^{\rm SM}_{\rm PEN} \label{eq:aaH}\\ 
	{\cal M}^{\rm BSM}(gg \to \gamma Z H) &=& \kappa_t {\cal M}^{\rm SM}_{\rm PEN} + \kappa_V {\cal M}^{\rm SM}_{\rm BX_1} \label{eq:aZH}\\
	{\cal M}^{\rm BSM}(gg \to Z Z H) &=& \kappa_t {\cal M}^{\rm SM}_{\rm PEN} + \kappa_V {\cal M}^{\rm SM}_{\rm BX_1} +
	\kappa_t^2\kappa_V {\cal M}^{\rm SM}_{\rm BX_2} + \nonumber \\ 
	&& \kappa_t\kappa_V\kappa_\lambda {\cal M}^{\rm SM}_{\rm TR_1} + 
	\kappa_t\kappa_V {\cal M}^{\rm SM}_{\rm TR_2} + \kappa_t\kappa_V^2 {\cal M}^{\rm SM}_{\rm TR_3}  \label{eq:ZZH}\\
	{\cal M}^{\rm BSM}(gg \to WW H) &=& \kappa_t {\cal M}^{\rm SM}_{\rm PEN} + \kappa_V {\cal M}^{\rm SM}_{\rm BX_1} +
	\kappa_t^2\kappa_V {\cal M}^{\rm SM}_{\rm BX_2} +   \nonumber \\
	&&  \kappa_t {\cal M}^{\rm SM}_{\rm BX_3} + \kappa_t\kappa_V\kappa_\lambda {\cal M}^{\rm SM}_{\rm TR_1} + 
	\kappa_t\kappa_V {\cal M}^{\rm SM}_{\rm TR_2} +  \nonumber \\
	&& \kappa_t\kappa_V^2 {\cal M}^{\rm SM}_{\rm TR_3} + \kappa_V {\cal M}^{\rm SM}_{\rm TR_4} \label{eq:WWH}
\end{eqnarray}
In the above, the amplitude  ${\cal M}^{\rm SM}_i$ is related to one of the diagram classes displayed in Fig.~\ref{fig:feyn-A-Z-H-pen-bx-tr} (Fig.~\ref{fig:feyn-WWH-pen-bx-tr} for $WWH$).
This can be easily identified by looking at $\kappa$-factors in front of the amplitude. Note that in $WWH$ amplitude, 
${\cal M}^{\rm SM}_{\rm TR_4}$ includes both (h) and (i) type diagrams of Fig.~\ref{fig:feyn-WWH-pen-bx-tr}.
This parametrization does not affect the gauge invariance of the amplitudes with respect to the gluons as 
it will become clear in the next section. The standard model prediction can be obtained by setting 
$\kappa_t=\kappa_V=\kappa_\lambda = 1$. Thus, except in $gg \to \gamma\gamma H$, we can expect nontrivial interference effects on total and differential cross sections for $gg \to VVH$ processes due to new physics 
in $\kappa$-framework.

\section{Calculation and Checks}
\label{sec:calc-check}

The calculation of quark-loop diagrams is carried out using a semi-automated in-house package {\tt OVReduce}~\cite{Agrawal:1998ch} which allows 
the calculation of any one-loop amplitude with maximum five propagators in the loop. The main steps involved in our calculation are: quark-loop trace evaluation, one-loop tensor reduction to master integrals and evaluation of master integrals. Trace calculation and 
simplification of the amplitude are done using symbolic manipulation software {\tt FORM}~\cite{Vermaseren:2000nd}. Tensor reduction 
of one-loop amplitudes into one-loop master integrals is done numerically following the method of Oldenborgh-Vermaseren~\cite{vanOldenborgh:1989wn}. 
Further, the one-loop master integrals are also calculated numerically using the {\tt OneLOop} package~\cite{vanHameren:2010cp}. 
More details on this can be found in ~\cite{Shivaji:2013cca}.
We perform the calculation in $4-2\epsilon$ space-time dimensions to regulate ultraviolet (UV) and infrared (IR) 
singularities of one-loop 
master integrals. Since the couplings of $Z$ and $W$ bosons with quarks involve $\gamma_5$, the trace calculation 
 needs special care. We have used 4-dimensional properties of $\gamma_5$ in the calculation. This works because the SM is anomaly free. We have chosen Unitary gauge for the calculation 
of the amplitudes.
\\

The amplitude calculation for each process can be efficiently organized using prototype 
amplitudes for each class of diagrams. For example, amplitudes for all the 12 independent pentagon diagrams in $gg \to \gamma\gamma H$ process can be obtained using only one prototype pentagon amplitude. 
Similarly, prototype amplitudes can be identified for each topology contributing to each process. 
The full amplitude for each process is a function of external momenta and polarization vectors/helicities.  
Due to huge expressions of the amplitudes, we calculate helicity amplitudes and the squaring of the 
amplitude for each process is done numerically. The number of helicity amplitudes for 
$gg \to \gamma\gamma H,~\gamma ZH,~ ZZH,~ WWH$ processes are 16, 24, 36, and 36, respectively. 
\\

There are a number of checks that we have performed in order to ensure the correctness of the amplitudes. We have checked 
that the amplitudes are separately UV and IR finite. In $4-2\epsilon$ dimensions, 
these divergences appear as poles in $1/\epsilon$ (for UV and IR) and $1/\epsilon^2$ (for IR only). Each pentagon 
diagram is UV finite. This we 
expect from the naive power counting. The individual box diagram is not UV finite, however, the full box 
amplitude, in each class, is UV finite. The UV finiteness of triangle amplitudes holds for each diagram.
One-loop diagrams with all massive internal lines are IR finite, as expected. Thus, IR finiteness check is relevant 
to the diagrams with massless quarks in the loop. This includes box class of diagrams of Fig.~\ref{fig:feyn-A-Z-H-pen-bx-tr}(b) in $gg \to \gamma ZH$ 
and $ZZH$. In $gg \to WWH$ case, potentially IR divergent diagrams include Fig.~\ref{fig:feyn-WWH-pen-bx-tr}(a), (b), (h) and (i).  
Unlike UV, the IR finiteness holds for each diagram~\cite{Shivaji:2013cca}.  
\\

We have also checked the gauge invariance of the amplitudes 
with respect to the external gluons. For that we numerically replace the 
gluon polarization vector $\epsilon^\mu(p)$ by its four momentum $p^\mu$ and 
expect a gauge invariant amplitude to vanish.
We find that the gauge invariance check holds for each class of diagrams. 
This is expected because different box and triangle topologies for each process arise due to 
the existence of various electroweak couplings.
This is a very strong check on our calculation which is organized using
only prototype amplitudes. However, this check 
cannot verify relative signs between different classes of diagrams. In order to verify such 
relative signs, one needs to perform gauge invariance check in electroweak theory 
which is a non-trivial task\footnote{A wrong relative sign between different class of diagrams {\it may} lead to violation of unitarity in certain 
processes~\cite{Maltoni:2001hu}.}. We rather rely on cross-checking the calculation 
using different methods and tools. 
We have compared our matrix element for each process with those calculated using 
{\tt MadLoop}~\cite{Alwall:2014hca} and have found an excellent agreement.
{Being process specific, our code is efficient and provides greater flexibility when producing phenomenological result.}
\\

Numerical predictions for cross section and kinematic distributions are obtained 
using Monte Carlo techniques for phase space integration. We use {\tt AMCI}~\cite{Veseli:1997hr} 
package for Monte Carlo phase space integration which is based on {\tt VEGAS}~\cite{Lepage:123074}
algorithm and allows parallelization of phase space point generation and matrix-element 
computation using {\tt PVM} software~\cite{Geist:1995:PPV:207505}.

\section{Numerical Results}\label{sec:numrRes} 

The cross section and kinematic distributions for $pp \to VVH$ processes in SM and in BSM constitute 
the main results of this section. 
The numerical results are produced using following basic selection cuts unless stated otherwise,
\begin{eqnarray}
p_T^{\gamma} > 50~{\rm GeV},~	|\eta^\gamma| < 2.5,~ \Delta R_{\gamma\gamma} > 0.4,~ |y^{H,Z,W}| < 5.
\label{eq:cuts}
\end{eqnarray}

The results for the $gg$ channel processes are calculated using {\tt CT14LO}~\cite{Dulat:2015mca} parton distribution function (PDF)
and partonic center-of-mass energy $(\sqrt{\hat s})$ is chosen as common scale for renormalization ($\mu_R$) and 
factorization ($\mu_F$). The results are obtained  for three different choices of collider energies: $\sqrt{s} = 14, 27,$ 
and 100 TeV. From phenomenological point of view we will focus on 
$p_T(H)$ and $M(VV)$ distributions.
\\

We compare the $gg$ channel contribution to $pp \to VVH$ with contribution arising from the $qq$ channels. The 
$qq$ channel contribution at LO and NLO (QCD) is calculated  using {\tt MadGraph5\_aMC@NLO}~\cite{Alwall:2014hca} in five flavor scheme for all but WWH production. The $qq$ channel contribution to WWH production is instead calculated in four flavor scheme\footnote{For WWH production, currently {\tt MadGraph5\_aMC@NLO} cannot produce NLO correction to the $bb$ channel.}.
The LO $qq$ channel contributions are pure electroweak processes and they do not depend on $\alpha_s$.
For LO and NLO (QCD) results, we use {\tt CTEQ14LO} and {\tt CT14NLO} PDFs, respectively~\cite{Dulat:2015mca}. The scale choice is same as in the $gg$ channel calculation. 
In both $gg$ and $qq$ channel calculations, the 
scale uncertainties are estimated by varying $\mu_R$ and $\mu_F$ independently by a factor of two. We 
quote only minimum and maximum uncertainties thus obtained. 
\\

To quantify the relative importance of the $gg$ channel contribution 
in processes dominated by the $qq$ channel, we 
define following ratio,
\begin{eqnarray}
R_1 = \frac{\sigma^{\rm VVH, LO}_{gg}}{\sigma^{\rm VVH, NLO}_{qq}-\sigma^{\rm VVH, LO}_{qq}}.
\label{eq:R1}
\end{eqnarray}

This ratio compares the leading order $gg$ channel contribution with NLO QCD correction in the $qq$ channel. Recall that technically $gg$ channels contribute at NNLO. 
Similarly, at differential level we define another ratio, 
\begin{eqnarray}
R_2 = \frac{\frac{d\sigma}{dX}\bigr\rvert^{\rm VVH, LO}_{gg}}{
\frac{d\sigma}{dX}\bigr\rvert^{\rm VVH, NLO}_{qq}},
\label{eq:R2}
\end{eqnarray}
where, $X$ denotes a kinematic variable.
\\

As mentioned in section~\ref{sec:bsm-para}, 
the BSM effects are parametrized in terms of scale factors 
$\kappa_t$, $\kappa_V$ and $\kappa_\lambda$.
In order to compare their relative importance, we vary 
them independently by 10\% about their SM values. 
Further, we comment on the effect of $\kappa_\lambda$ 
and $\kappa_{HHVV}$ (the scale factor for the $HHVV$ coupling\footnote{Note this is different from $k_V$, which scales both $HVV$ and $HHVV$ couplings at the same time. }) 
which are least constrained at present,  
in $ZZH$ and $WWH$ processes.

\subsection{Predictions for the $pp \to \gamma \gamma H$ process}\label{sec:results-AAH}

\begin{table}[H]
 \begin{center}
 \resizebox{0.7\columnwidth}{!}{%
  \begin{tabular}{|c|c|c|c|c|c|c|c|c|}
   \hline
$\sqrt{\rm s}\;(\rm TeV)$ 
& {$\sigma^{\gamma\gamma\tin{\rm H,\,LO}}_{\tin{gg}}\;~[\rm ab]$} 
& {$\sigma^{\gamma\gamma\tin{\rm H,\,LO}}_{\tin{qq}}\;~[\rm ab]$} 
& {$\sigma^{\gamma\gamma\tin{\rm H,\,NLO}}_{\tin{qq}}\;[\rm ab]$} 
\\
   \hline
& {} & {} &   {}   \\
   
14   
& {$5.36^{+28\%}_{-20\%}$}  
& {$0.033^{+13\%}_{-14\%}$}  
& {$0.046^{+5\%}_{-6\%}$} 
\\
  
& {} & {} &   {}   \\
   
27   
& {$22.0^{+22\%}_{-19\%}$} 
& {$0.153^{+15\%}_{-17\%}$}  
& {$0.234^{+5\%}_{-7\%}$} 
\\

& {} & {} &   {}   \\
   
100   
& {$220.1^{+27\%}_{-21\%}$}  
& {$1.4^{+20\%}_{-20\%}$}  
& {$2.25^{+5\%}_{-8\%}$} 
\\
     & & &  \\ 
   \hline  \end{tabular}}
 \end{center}
 \caption{A comparison of different perturbative orders in QCD coupling contributing to $pp \to \gamma \gamma H$
hadronic cross section at $\sqrt{s}=$ 14, 27, and 100 TeV. } \label{table:xs-aaH}
\end{table}

The cross section for this process is dominated by the $gg$ channel. In the $qq$ channel, only bottom-quark initiated 
subprocess contribute to $\gamma\gamma H$ production. 
However, this  cross section is quite small, owing to small 
bottom Yukawa coupling. In Tab.~\ref{table:xs-aaH}, we compare the $gg$ and $qq$ channel
contributions to the hadronic cross section at 14, 27 and 100 TeV colliders. The results are with minimum 50 GeV transverse momentum of photons. We find that the $gg$ channel contribution increases 40 times as the collider center-of-mass energy goes
from 14 TeV to 100 TeV. Due to a small cross-section, this process cannot be observed at the HL-LHC; FCC-pp will be more suitable.
The $gg$ channel contribution becomes important at higher center-of-mass energy collider, as in this case smaller partonic momentum fractions ($x$) are accessible, where gluon flux is significantly large. 
The scale uncertainties on the
cross sections for the $gg$ channel are in the range of 20-30\%. It is clear 
from the table that the $qq$ channel contribution is negligible compared to 
the $gg$ channel contribution. It is merely 1\% of the $gg$ channel contribution even after 
including the NLO-QCD corrections.
\\

\begin{figure}
\begin{center}
\includegraphics[angle=0,width=0.48\linewidth]{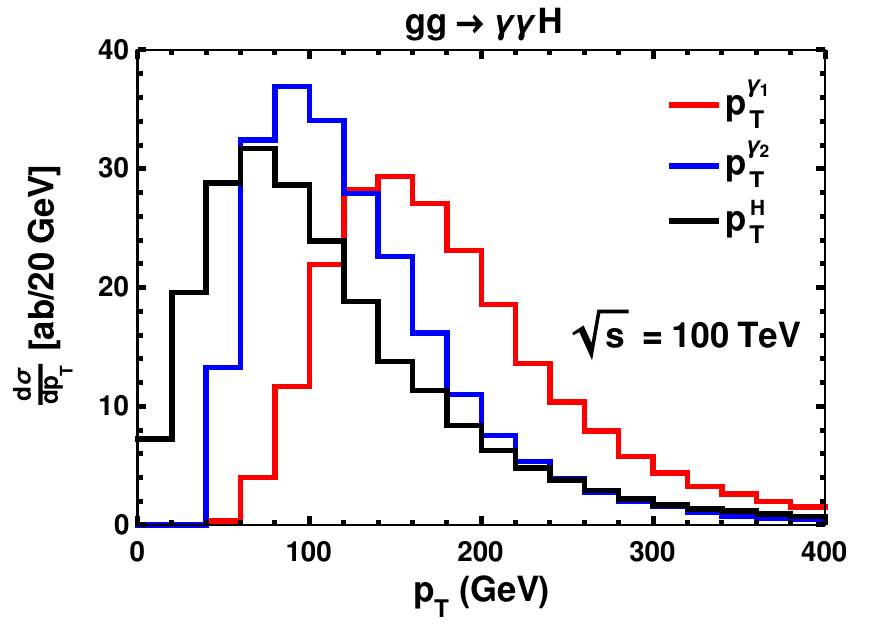}
\includegraphics[angle=0,width=0.48\linewidth]{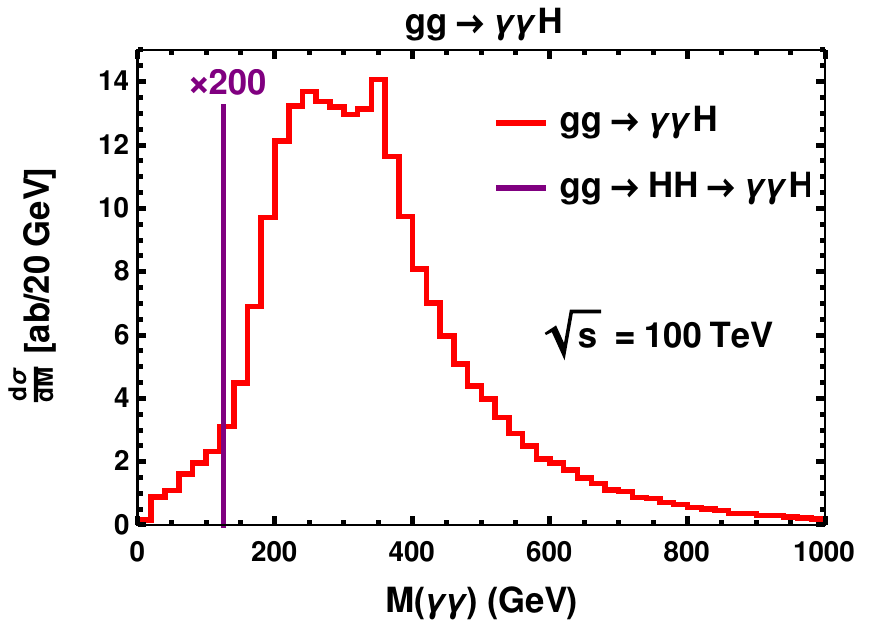}
\end{center}
\caption{ Kinematic distributions for $\GGAAH$ process in the SM at 100 TeV. In the $p_T$ distribution plot, $\gamma_1$ and $\gamma_2$ refer to the hardest and second hardest photons in $p_T$, respectively.  In the right plot, we show $M(\gamma \gamma)$ distribution for $gg \to \gamma\gamma H$. In addition, the total cross section for the $gg \to H H \to \gamma\gamma H$ process has been shown at 125 GeV. ``$\times 200$" implies that the height of purple vertical line should be multiplied by a factor of 200 in order to get the correct cross section for the $gg \to H H \to \gamma\gamma H$ process. } 
\label{fig:dists-AAH-sm}
\end{figure}

In Fig.~\ref{fig:dists-AAH-sm}, we have plotted $p_T$ distributions for hardest photon, next-to-hardest photon, and Higgs in the left figure, and 
diphoton invariant mass distribution ($M(\gamma\gamma)$) in the right figure for the 100 TeV collider (FCC-hh). The $ p_T$ distributions for them peak around 150 GeV, 90 GeV, and 70 GeV, respectively. 
We find that the tail of $p_T(H)$ is softer than that of photons. The $M(\gamma\gamma)$ distribution shows an interesting feature -- it has two peaks. The right peak occurs at around 350 GeV, exhibiting the $t\bar{t}$ threshold effect in the distribution. To verify that the second peak is indeed due to $t\bar{t}$ threshold effect, we changed in our code the  value of $m_t$ from 173 GeV to 200 GeV and the second peak was found to get shifted to 400 GeV.
\\

As mentioned before, this process is a background to double Higgs production process when one of the Higgs bosons decays into a photon pair. To manage the background one usually looks at `$\gamma \gamma b \bar{b}$' final state, instead of `$ b \bar{b} b \bar{b}$', as the signature of the double Higgs boson production.  At a 100 TeV collider,  while the cross section for the $gg \to \gamma \gamma H$ production, with the cuts in Eq. \ref{eq:cuts}, is about 220 ab, the cross section for $gg \to H H \to \gamma\gamma H$, with the same set of cuts, is about 2600 ab.
From the right panel of Fig.~\ref{fig:dists-AAH-sm}, it can be seen that the cross section for $\gamma\gamma H$ production in the bin from 120 GeV to 140 GeV is about 3 ab. On the other hand, all the cross section for $gg \to H H \to \gamma\gamma H$ is concentrated in a very narrow width around the mass of Higgs, 125 GeV \footnote{In the right panel of Fig.~\ref{fig:dists-AAH-sm}, at 125 GeV, rather than showing a very narrow Breit-Wigner distribution, we have shown the total cross section for $gg \to H H \to \gamma\gamma H$ by a single vertical line.}. As a result, $gg \to \gamma\gamma H$ is an insignificant background to the process $gg \to H H \to \gamma\gamma H$.
\\

Regarding anomalous coupling contributions, we note that  as only pentagon diagrams contribute to the process $gg \to \gamma\gamma H$, its cross section scales as $\kappa_t^2$. So a 10\% change in $\kappa_t$ will change the cross section and
distributions by about 20\%. For the $qq$ channel process, the cross section is too small. It depends on $\kappa_b$, which we do not change from the standard model value.

\subsection{Predictions for the $pp \to \gamma ZH$ process}\label{sec:results-AZH}

Unlike $\gamma\gamma H$ case,  the $\gamma ZH$ production receives dominant  contribution from the $qq$ channel.  
With $ p_T^\gamma > 50$ GeV, the $gg$ channel contributions to $\gamma ZH$ production at 14, 27, and 100 TeV colliders are 4 ab, 16 ab, and 168 ab, respectively. 
The corresponding values for the LO $qq$ channel contribution are 689 ab, 1733 ab, and 7498 ab, respectively.
From Tab.~\ref{table:xs-aZH}, it can be seen that $R_1$, which is the ratio of the $gg$ channel contribution to NLO correction in the $qq$ channel, is as small as 0.06 for 100 TeV collider, and even smaller for HE-LHC (27 TeV) and LHC (14 TeV).  
The scale uncertainties for the $gg$ channel are around 20\% while those for the $qq$ channel at NLO are  in the range of $2-3\%$. A larger 
scale dependence in the $gg$ channel contribution can be attributed to the presence of higher power of $\alpha_s$ factor in the $gg$ amplitudes.

\begin{table}[H]
 \begin{center}
  \resizebox{0.8\columnwidth}{!}{%
  \begin{tabular}{|c|c|c|c|c|}
   \hline
   $\sqrt{\rm s}\;(\rm TeV)$ & \multicolumn{1}{c|}{$\sigma^{\gamma\tin{\rm ZH,\,LO}}_{\tin{gg}}\;~[\rm ab]$} & \multicolumn{1}{c|}{$\sigma^{\gamma\tin{\rm ZH,\,LO}}_{\tin{qq}}\;~[\rm ab]$} & \multicolumn{1}{c|}{$\sigma^{\gamma\tin{\rm ZH,\,NLO}}_{\tin{qq}}\;[\rm ab]$} & \multicolumn{1}{c|}{$R_1$}\\
\cline{2-5}
   \hline
   & \multicolumn{1}{c|}{} &\multicolumn{1}{c|}{}  &\multicolumn{1}{c|}{} & \multicolumn{1}{c|}{}   \\

  14   & \multicolumn{1}{c|}{$4.0^{+26\%}_{-20\%}$} & \multicolumn{1}{c|}{$689^{\,+\, 0\%}_{-0.2\%}$}  &  \multicolumn{1}{c|}{$909^{+1.7\%}_{-1.3\%}$} &   \multicolumn{1}{c|}{0.02} \\

   & \multicolumn{1}{c|}{} &\multicolumn{1}{c|}{}  &\multicolumn{1}{c|}{} & \multicolumn{1}{c|}{}   \\

  27   & \multicolumn{1}{c|}{$16^{+22\%}_{-17\%}$} & \multicolumn{1}{c|}{$1773^{+3.0\%}_{-3.6\%}$} &  \multicolumn{1}{c|}{$2349^{+1.7\%}_{-2.1\%}$}  &  \multicolumn{1}{c|}{0.03}  \\

   & \multicolumn{1}{c|}{} &\multicolumn{1}{c|}{}  &\multicolumn{1}{c|}{} & \multicolumn{1}{c|}{}   \\

  100   & \multicolumn{1}{c|}{$168^{+21\%}_{-19\%}$} & \multicolumn{1}{c|}{$7498^{+8.8\%}_{-9.4\%}$}  &  \multicolumn{1}{c|}{$10430^{+2.2\%}_{-3.8\%}$}  &  \multicolumn{1}{c|}{0.06}\\
     & & & & \\   
   \hline  \end{tabular}}
 \end{center}

 \caption{ A comparison of different perturbative orders in QCD coupling contributing to $pp \to \gamma ZH$
hadronic cross section at $\sqrt{s}=$ 14, 27, and 100 TeV. $R_1$ compares 
the $gg$ channel contribution with correction at NLO and it is defined in Eq~\ref{eq:M1}.} \label{table:xs-aZH}
\end{table}

In Tab.~\ref{table:xs-aZH-cuts}, the effect of various $p_T^\gamma$ cuts in $gg$ and $qq$ channels has been shown. As the cut on $p_T^\gamma$ increases, the $qq$ channel cross section decreases  faster than the $gg$ channel. In going from 50 GeV to 200 GeV cut, the cross section of the $gg$ channel decreases roughly by a factor of 6, while that of the $qq$ channel decreases by a factor of 9. Thus relative 
contribution from the $gg$ channel can be enhanced with the help of harder 
$p_T^\gamma$ cut. We find that the $p_T(H)$ cuts have opposite effect \emph{i.e.}
the $gg$ channel is favored at low $p_T(H)$.

\begin{table}[H]
\begin{center}
\resizebox{0.95\columnwidth}{!}{%
\begin{tabular}{|c|c|c|c|}
\hline
$p_{T, min}^\gamma$ (GeV) &  ${gg \rightarrow \gamma \rm Z H }$ [ab] &  ${qq \rightarrow \gamma \rm Z H (LO)}$ [ab] & ${qq \rightarrow \gamma \rm Z H (NLO)}$ [ab] \\
\hline
50 & 168 & 7498 & 10430 \\
\hline
100 & 95 & 2812 & 4072 \\
\hline
150 & 47 & 1366 & 2069 \\
\hline
200 & 28 & 765  & 1190 \\
\hline
\end{tabular}
}
\end{center}
	\caption{Effect of $p_T^\gamma$ cut on the cross section of $pp \rightarrow \gamma Z H$ production at the 100 TeV collider (FCC-hh).}
\label{table:xs-aZH-cuts}
\end{table}

\begin{figure}[H]
\begin{center}
\includegraphics[angle=0,width=0.48\linewidth]{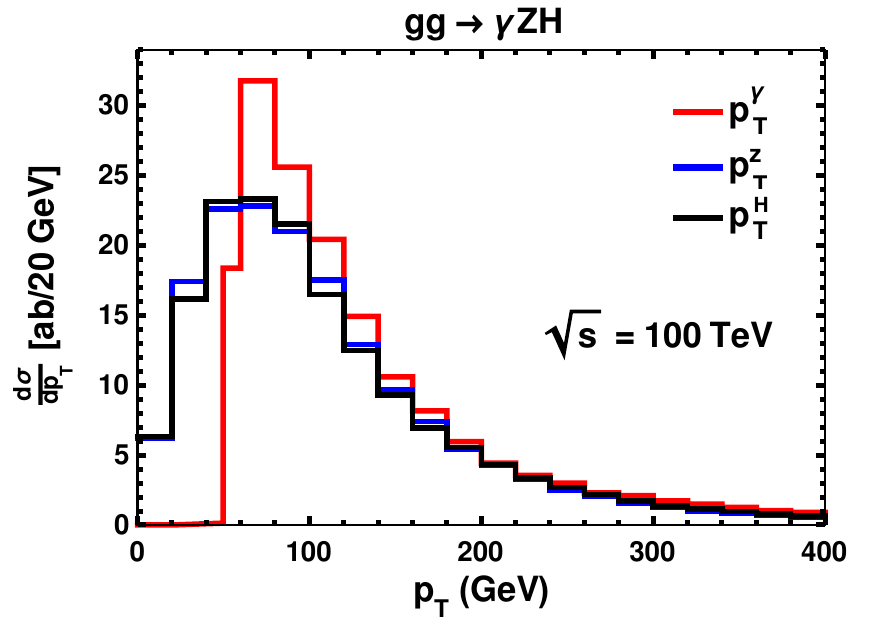}
\includegraphics[angle=0,width=0.48\linewidth]{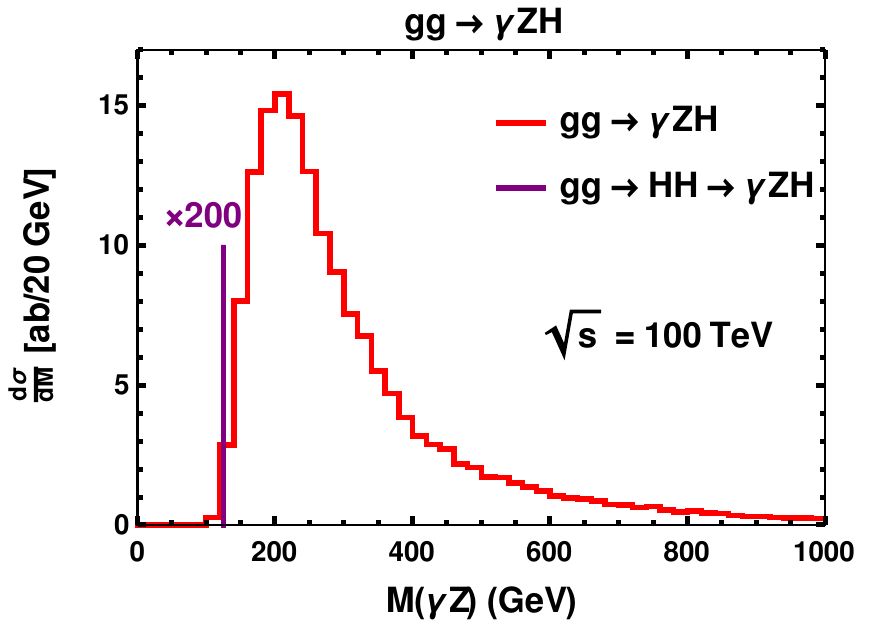}
\end{center}
\caption{ Kinematic distributions for $\GGAZH$ in the SM at 100 TeV. The purple vertical line in the right plot at 125 GeV shows the total cross section for the process $gg \to H H \to \gamma Z H$. ``$\times 200$" means that the height of the purple vertical line needs to be scaled by a factor of 200 to get the correct cross section for the $gg \to H H \to \gamma Z H$ process. }
\label{fig:dists-AZH-sm}
\end{figure}

In Fig.~\ref{fig:dists-AZH-sm}, we have displayed $p_T$ distributions for the final state particles on the left, and $\gamma Z$ pair invariant mass distribution on the right for the 100 TeV collider. The $p_T$ distributions peak around  100 GeV while the $M(\gamma Z)$ distribution peaks around 200 GeV.  Like the  case of $gg \to \gamma \gamma H$ process as a background to $gg \to H H \to \gamma \gamma H$, the $gg \to \gamma  Z H$ process is also an insignificant background to $gg \to H H \to \gamma Z H$. This is because at a 100 TeV collider, with the cuts in Eq.~\ref{eq:cuts}, the cross section for $gg \to H H \to \gamma Z H$ is about 2000 ab, while 
 the cross section for $gg \to \gamma  Z H$ process is about 170 ab. Moreover, all the cross section for the $gg \to H H \to \gamma Z H$ process congregates around the  mass of the decaying Higgs boson, 125 GeV
\footnote{However, instead of showing a very narrow Breit-Wigner distribution for Higgs' decay, we have depicted the total cross section at 125 GeV by a single vertical line.}, while, as can be seen from the right panel of the Fig.~\ref{fig:dists-AZH-sm}, the cross section for the $gg \to \gamma  Z H$ process in the bin from 120 GeV to 140 GeV  is about 3 ab. However, the $qq$ channel for $\gamma Z H$ production may act as an important background for the $gg \to H H \to \gamma Z H$ process.
\\

\begin{figure}[H]
\begin{center}
\includegraphics[angle=0,width=0.48\linewidth]{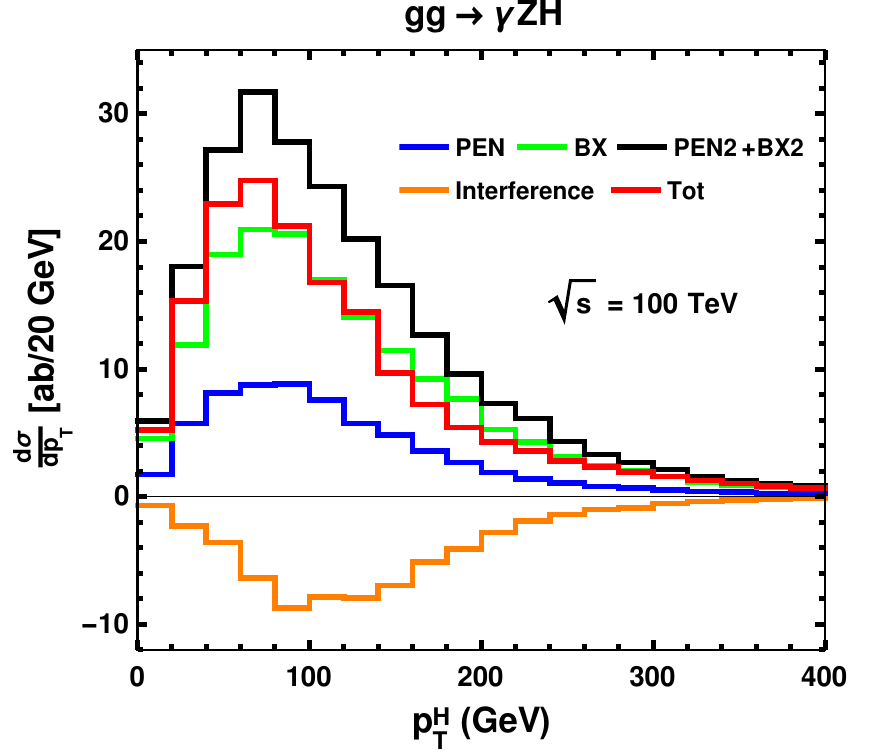}
\includegraphics[angle=0,width=0.48\linewidth]{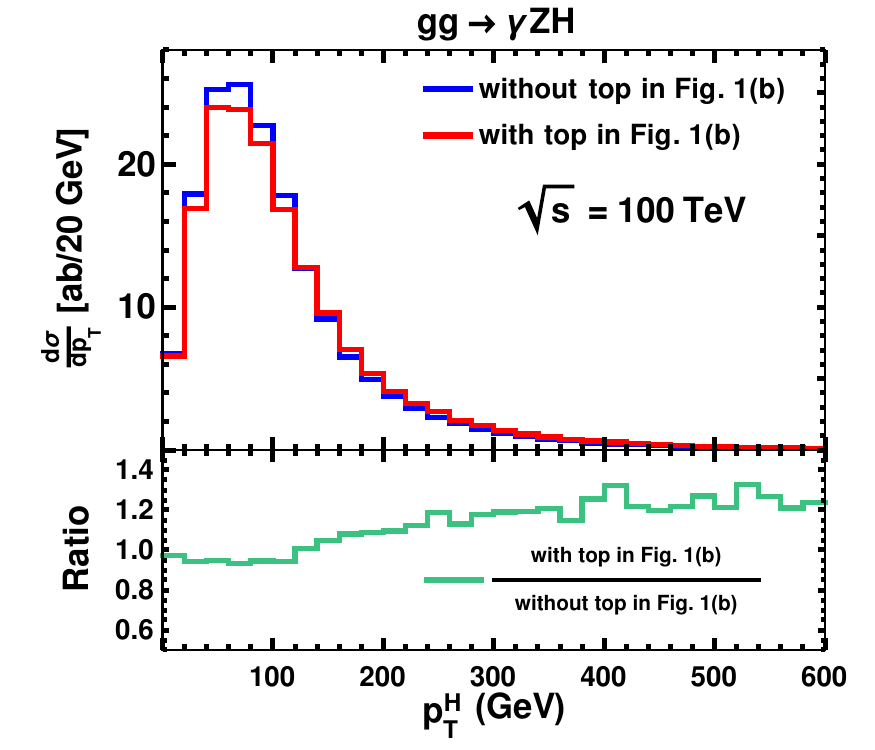}
\end{center}
	\caption{Left: The contribution of pentagon (blue) and box (green) diagrams, as well as their squared sum, interference, and total contribution  to $p_T(H)$ distributions for the $\GGAZH$ process at the 100 TeV FCC-hh collider. Right: The effect of excluding top-quark contribution from the diagrams in Fig. 1(b) to the full amplitude.} \label{fig:dist-aZH-pen-bx-tr}
\end{figure}

In Fig.~\ref{fig:dist-aZH-pen-bx-tr},  we show  $ p_T(H)$ distributions for different classes of diagrams -- pentagon, box, and sum of their individual contributions, their interference, and total at the 100 TeV collider. 
The contribution of the box diagrams is more than the pentagon diagrams mainly because of the light quark contributions. 
The interference effect between the pentagon and box diagrams has kinematic dependence. 
We find that in the region of our kinematic interest, it is always destructive and, near the peak, its effect is close to -30\%.
\\

Since the $gg\gamma Z^*$ type box amplitude does 
 not depend on the axial-vector coupling of the off-shell longitudinal $Z$-boson with the quarks, the top-quark contribution  is not very significant at the level of total cross section. This is shown in the right panel of the Fig.~\ref{fig:dist-aZH-pen-bx-tr}. We can see that in the 
tail where top quark is effectively light, the cross section increases by about 20\%. 
\\


\begin{figure}[H]
\begin{center}
\includegraphics[angle=0,width=0.48\linewidth]{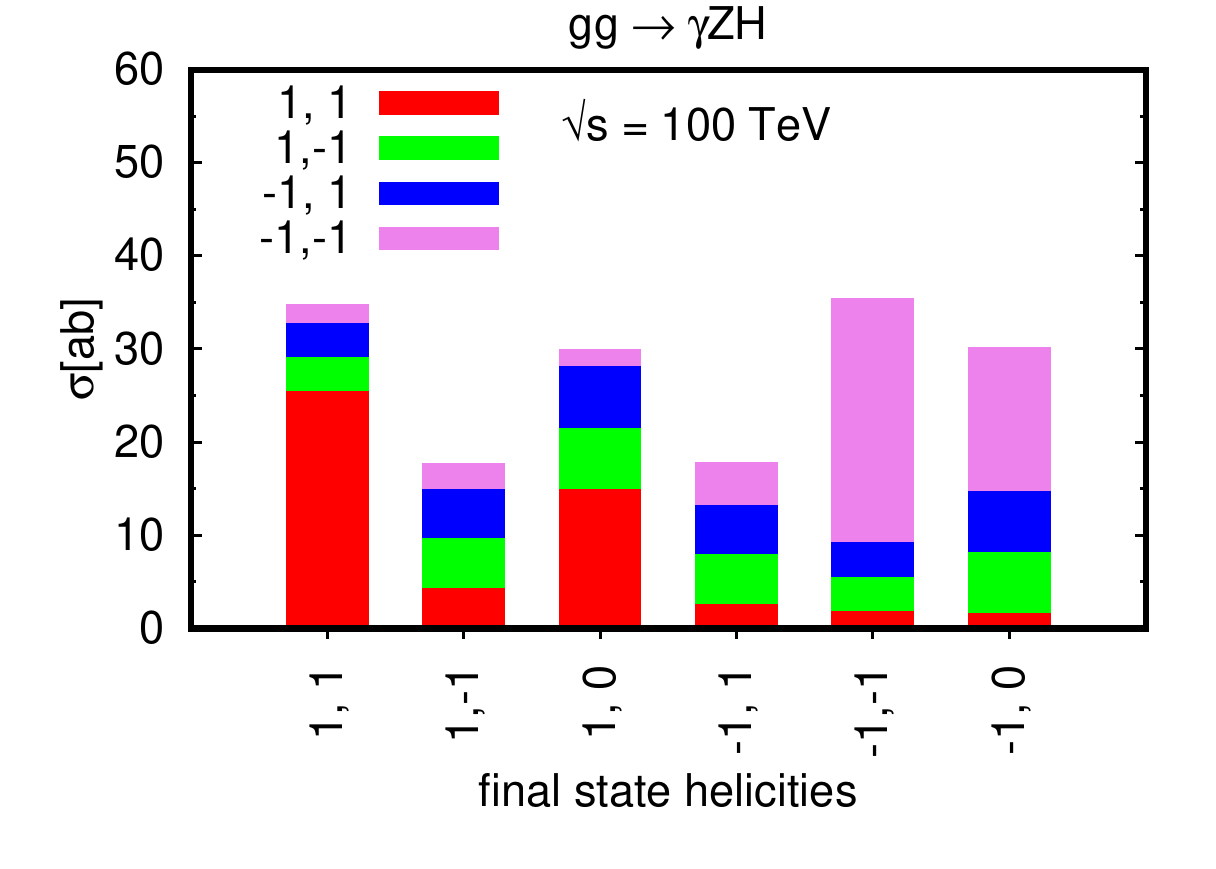}
\includegraphics[angle=0,width=0.48\linewidth]{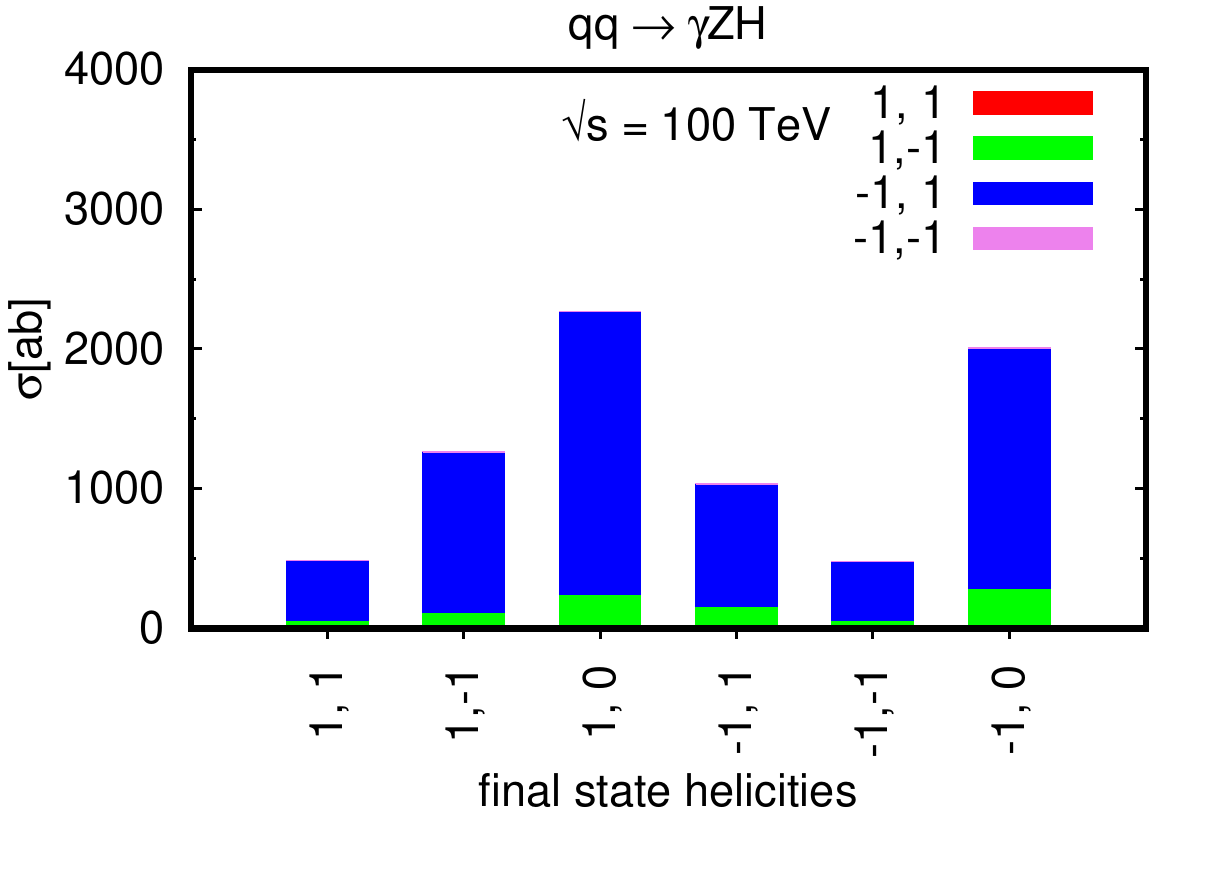}
\end{center}
\caption{ LO cross section for $\gamma ZH$ production in different helicity configurations in the $gg$ (left) and $qq$ (right) channels. Legends correspond to different helicities of initial states.}\label{fig:xs-hist-helicity-aZH}
\end{figure}

\begin{figure}[H]
\begin{center}
\includegraphics[angle=0,width=0.48\linewidth]{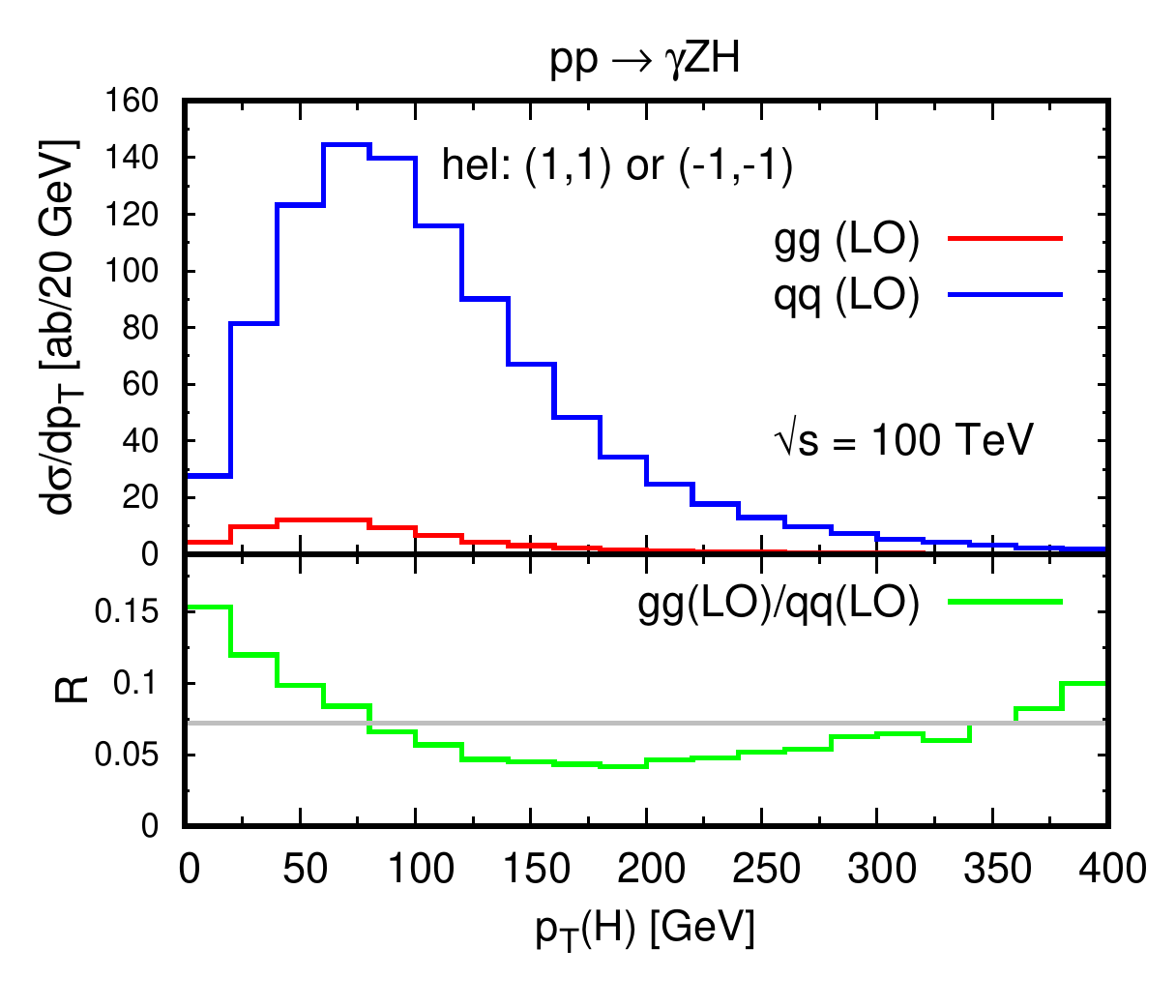}
\includegraphics[angle=0,width=0.48\linewidth]{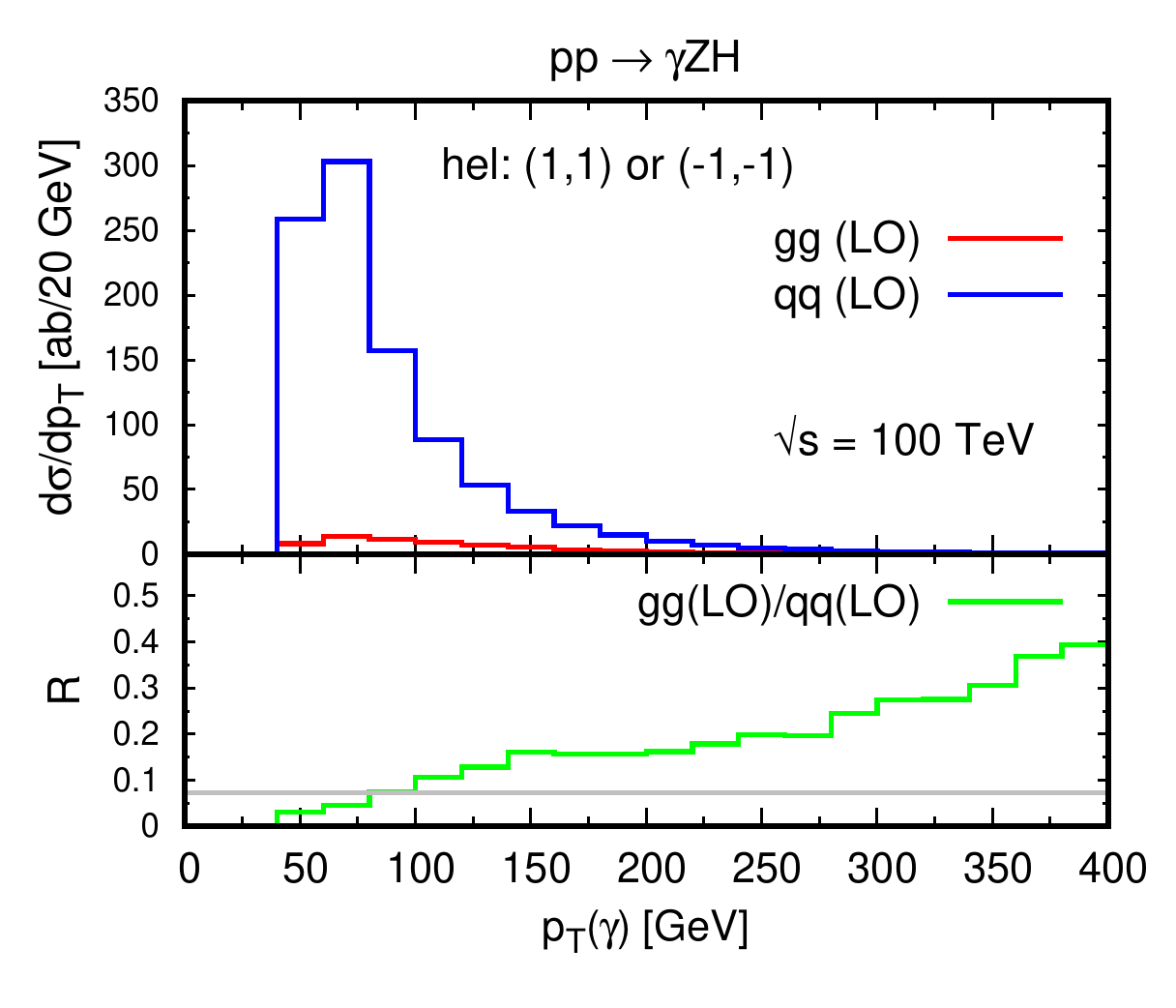}
\end{center}
\caption{The kinematic distributions for $gg$ and $qq$ channels when both $\gamma$  and $Z$-boson have same helicity. The ratio of the distributions from the two channels is shown in the lower panel of each figure.}\label{fig:dist-helicity-aZH}
\end{figure}

We have noted that the relative importance of gluon fusion 
channel can be enhanced by applying higher $p_T(\gamma)$ cuts.  
To distinguish the $gg$ channel contribution from the dominant $qq$ channel, one can use the polarized cross sections and distributions. In Fig. \ref{fig:xs-hist-helicity-aZH}, we have displayed the LO cross sections  for various  helicity states of the final state
particles, $\gamma$ and $Z$ boson. The figure also shows the contribution of
various polarization states of the initial state particles. We cannot measure the polarization of the initial
state particles that are in a bound state, proton. However, experimentally, one can
measure the  $Z$-boson polarization~\cite{Chatrchyan:2011ig,Aad:2012ky,Aaboud:2019gxl}. The polarization of photon has been measured by the LHCb collaboration in $b$-baryon's decay\cite{Aaij:2016ofv,Legger:2006cq, Hiller:2007ur, Orlovsky:2007hv, Shchutska:2008dba, Oliver:2010im, Martin:2019xcc}. At a 100 TeV collider,
the contribution of the $gg$ channel process to the production of
$\gamma Z H$ is only $2.2\%$. However, if we look at those final states where
photon and $Z$-boson have the same transverse  polarization, then this ratio increases to $10-11\%$. (The $qq$ channel makes largest contribution when the $Z$ boson is longitudinally polarized.)
This is a non-trivial contribution, and can be measured, if enough integrated luminosity
is available. In Fig.\ref{fig:dist-helicity-aZH}, we have plotted the Higgs boson  and $Z$-boson $p_T$ distributions. By making appropriate cuts on the small and large  $p_T$
of these particles, we can further enhance the $gg$ channel contribution.
\\


Turning to the effect of anomalous couplings, we find that the $gg$ channel shows very small dependence on the $\kappa_t$, as it is present only in pentagon diagrams whose contribution is small (see Fig.~\ref{fig:dist-aZH-pen-bx-tr}). However, it strongly depends on $\kappa_V$, as the box-diagrams contribution is much more than the pentagon-diagram  contribution. 
We find that the change in cross section for $\kappa_t= 1.1 (0.9)$ is 
5.4\% (-1.2\%). On the other hand, for $\kappa_V= 1.1 (0.9)$ the 
cross section changes by 18\% (15\%).
{ We do not show the effect of anomalous couplings on the distribution. It can be understood qualitatively from  Eq.~\ref{eq:aZH} and Fig.~\ref{fig:dist-aZH-pen-bx-tr} in the $gg$ channel}. 
The $qq$ channel is sensitive to $\kappa_V$ only. The amplitude has
overall linear dependence on $\kappa_V$ due to which the effect of anomalous coupling 
$k_V$ is flat for both total and differential cross sections.


\subsection{Predictions for $pp \to ZZH$}


The cross sections for $ZZH$ production via various channels have been tabulated in Table~\ref{table:xs-ZZH} along with the corresponding scale uncertainties.  The total cross section for $gg \to ZZH$ is significantly larger than that of $gg \to \gamma Z H$. This increase is mainly due to the 
contribution from axial-vector coupling of $Z$ with quarks. 
The $gg$ channel contributions to $ZZH$ production at 14, 27, and 100 TeV colliders are 124 ab, 579 ab, and 7408 ab, respectively. The corresponding values of the LO $qq$ channel contributions are 2184, 5997, and 36830 ab, respectively. The ratio, $R_1$, is found to be 0.25, 0.4, and  1.05, respectively. 
Thus at 100 TeV, the $gg$ channel contribution is as important as the QCD NLO correction. As has already been discussed, this increase in ratio $R_1$ with collider energy is due to the large gluon flux. 
\\

\begin{table}[h]
 \begin{center}
  \resizebox{0.8\columnwidth}{!}{%
  \begin{tabular}{|c|c|c|c||c||}
   \hline
   $\sqrt{\rm s}\;(\rm TeV)$ & $\sigma^{\tin{ \rm ZZH,\,LO}}_{\tin{gg}}\;~[\rm ab]$ & $\sigma^{\tin{\rm ZZH,\,LO}}_{\tin{qq}}\;~[\rm ab]$ & $\sigma^{\tin{\rm ZZH,\,NLO}}_{\tin{qq}}\;[\rm ab]$ & ${R}_{1} $\\
   \hline
   & & & & \\
  14   & $124^{+28.2\%}_{-21.0\%}$ & $2184^{+0.2\%}_{-0.6\%}$ & $2710^{+1.4\%}_{-1.0\%}$ & 0.24  \\
   & & & & \\
  27   & $579^{+23.3\%}_{-18.5\%}$ & $5997^{+2.4\%}_{-3.0\%}$ & $7396^{+1.3\%}_{-1.6\%}$  & 0.41 \\
   & & & & \\
  100   & $7408^{+22\%}_{-18\%}$ &   $36830^{+8.0\%}_{-8.7\%}$  & $43940^{+1.2\%}_{-2.6\%}$ &  1.04 \\
     & & & & \\     
   \hline
  \end{tabular}}
 \end{center}
 \caption{ A comparison of different perturbative orders in QCD coupling contributing to $pp \to ZZH$  cross section at $\sqrt{s}=$ 14, 27, and 100 TeV. The ratio $R_1$, defined in Eq.~\ref{eq:R1}, quantifies the $gg$  channel contribution with respect to the NLO correction in $qq$ channel process. } \label{table:xs-ZZH}
\end{table}

In the $gg$ channel, the scale uncertainties of the total cross sections are in the range of 20-30\% which is similar to the scale uncertainties observed for $\gamma\gamma H$ and $\gamma ZH$. We find that the uncertainty due to the renormalization scale variation is more than that due to the factorization scale variation. While the change in the renormalization scale mainly changes $\alpha_s$, the change in the factorization scale changes the parton distribution function. 
The uncertainty for the renormalization scale variation is nearly same at all the collider energies. This happens as the contribution to the total cross section comes from nearly same region of partonic center of mass energy of the process and in every bin of this region, $\alpha_s$ changes by nearly same factor for the change in the renormalization scale. 
However, uncertainty for the factorization scale variation is different for different colliders.
This happens as for different collider energies, different $x$ regions contribute to the process and for different $x$ regions change in parton distribution function with the factorization scale is different, where $x$ is partonic momentum fraction. We have also observed that with an increase in the factorization scale, for 14 and 27 TeV colliders, the cross-section decreases; however for 100 TeV collider the cross-section increases.
\\

\begin{figure}[H]
\begin{center}
\includegraphics[angle=0,width=0.48\linewidth]{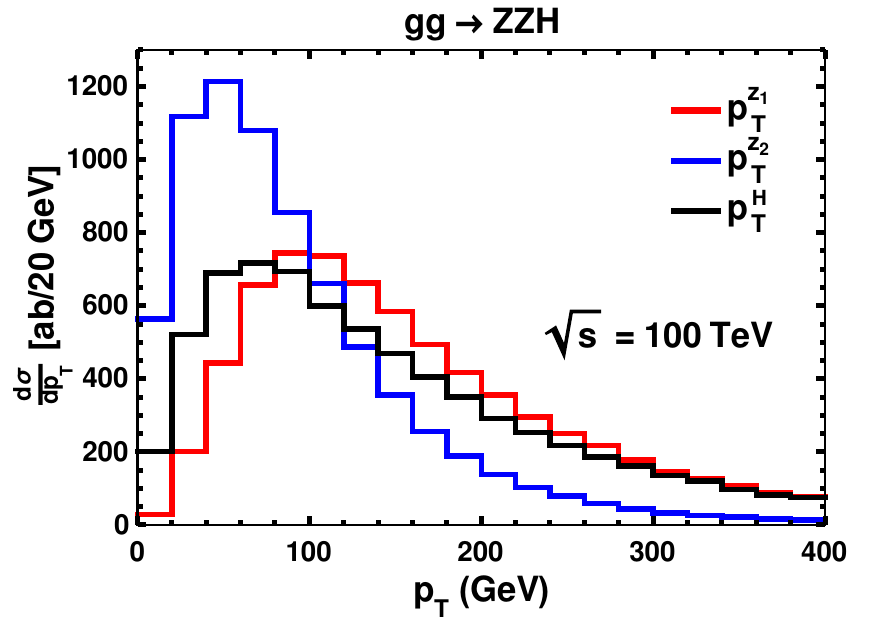}
\includegraphics[angle=0,width=0.48\linewidth]{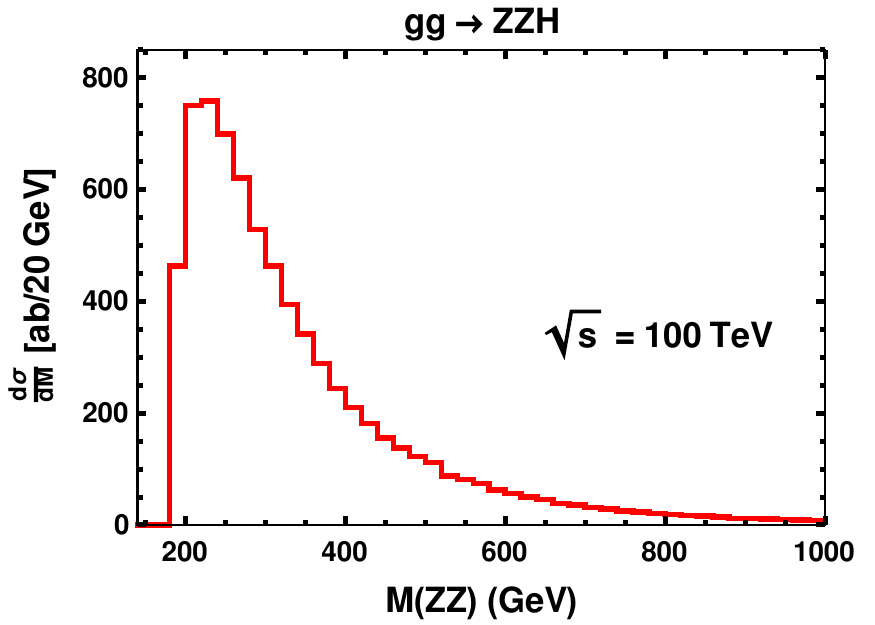}
\end{center}
	\caption{ Kinematic distributions for $\GGZZH$ in the SM at the 100 TeV collider.  $Z_1\ {\rm and}\ Z_2$ refer to the hardest, and second hardest in $p_T$, respectively. } 
\label{fig:dists-ZZH-sm}
\end{figure}

In the tree level $qq$ channel, there is no QCD vertex. So here change in the renormalization scale does not affect the cross section. But, the change in the factorization scale can affect the cross section, and uncertainty increases with collider energy. However, when  NLO QCD correction is considered, change in either of renormalization and factorization scales changes the cross section. The uncertainty in the cross section due to the renormalization scale variation is small as NLO QCD correction is much smaller than the LO results.
The overall uncertainty in this case is smaller than the LO case, which is expected for higher order calculation. 
\\

\begin{figure}[H]
\begin{center}
\includegraphics[angle=0,width=0.48\linewidth]{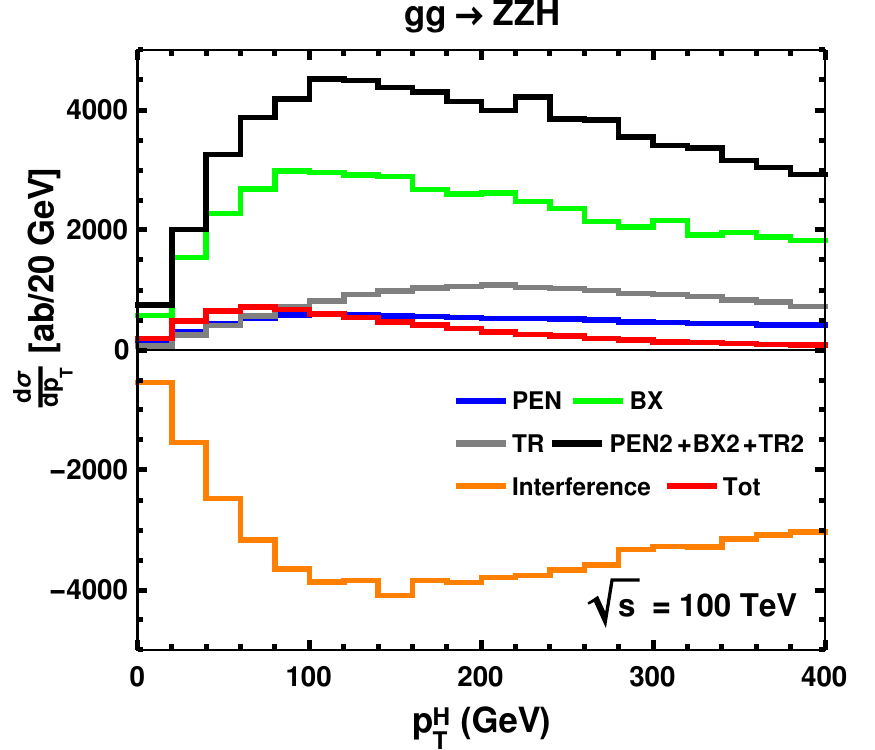}
\includegraphics[angle=0,width=0.48\linewidth]{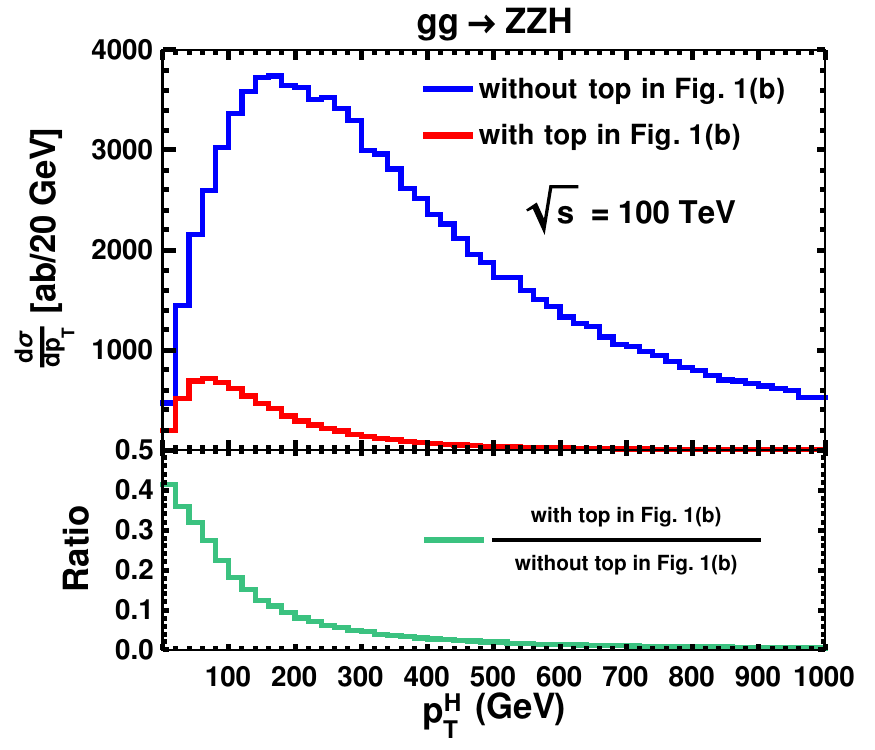}
\end{center}
\caption{Left : SM contribution of pentagon (blue), box (green), triangle (gray) diagrams, as well as their squared sum (black), interference (orange) and total (red) contribution to $p_T(H)$ distributions in $\GGZZH$ at 100 TeV collider (FCC-hh). 
Right: The effect of excluding top-quark contribution from Fig. 1(b) to full amplitude. } \label{fig:dist-ZZH-pen-bx-tr}
\end{figure}

\begin{figure}[h]
\begin{center}
\includegraphics[angle=0,width=0.48\linewidth,height=0.35\linewidth]{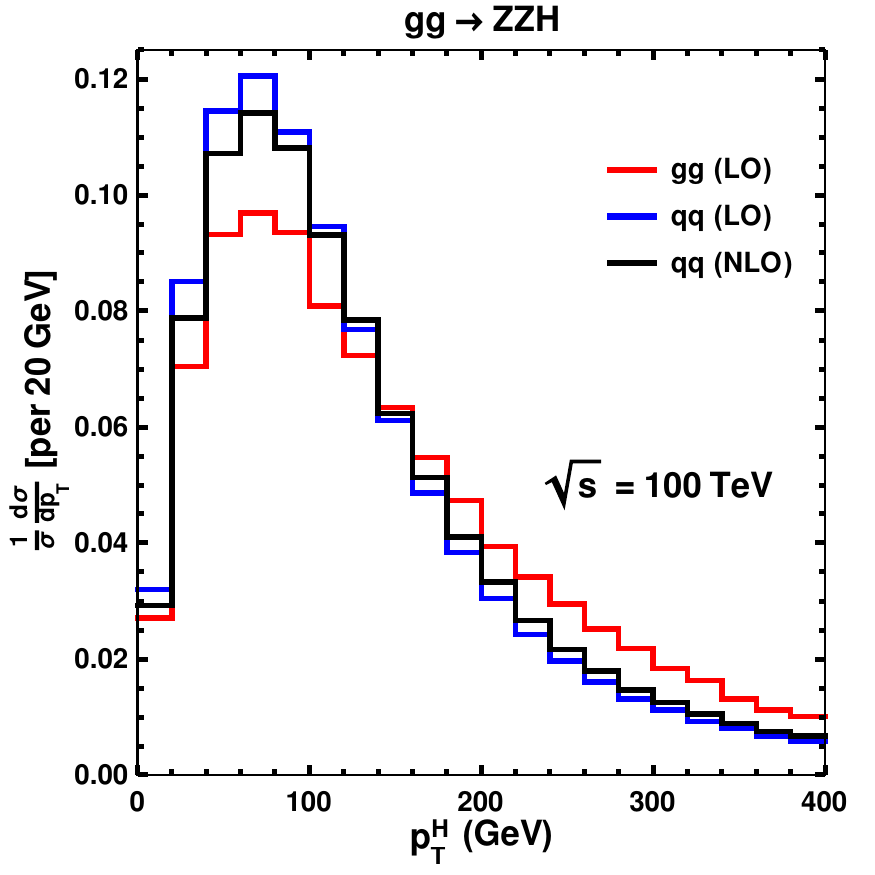}
\includegraphics[angle=0,width=0.48\linewidth,height=0.35\linewidth]{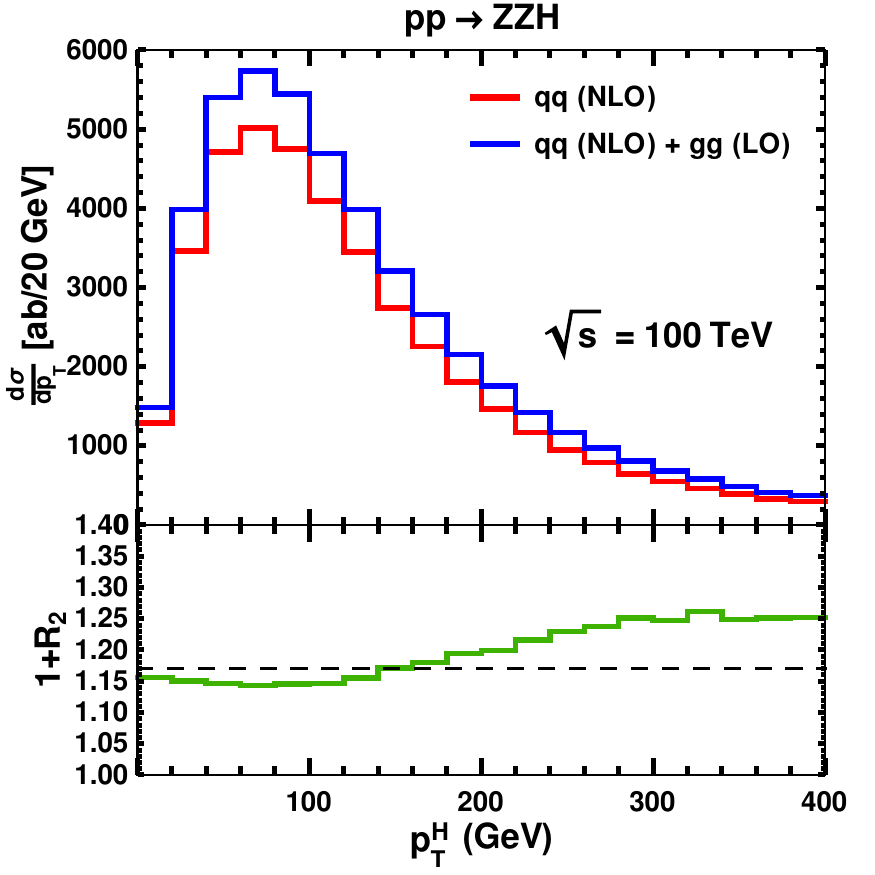}
\end{center}
	\caption{The left figure shows the normalized distribution for $p_T(H)$ in  $gg$ and $qq$ channel processes. In the top panel of the right figure, we show the distribution of $qq$ (NLO) + $gg$ (LO) and $qq$ (NLO) production with $p_T(H)$. The lower panel shows the ratio of them.
}  
 \label{fig:dist-ZZH-QQ-GG}
\end{figure}

In Fig.~\ref{fig:dists-ZZH-sm}, we have plotted $p_T$ distributions for leading $ p_T(Z_1)$, next-to-leading $ p_T(Z_2)$, and Higgs boson in the left figure, and $Z$-pair invariant mass distribution in the right figure for the 100 TeV collider. The $p_T$ distributions peak around 100 GeV, 60 GeV, and 80 GeV, respectively.  The $M(ZZ)$ distribution peaks around 
$Z$ pair threshold. 
\\

Interference of various diagrams plays a major role in $gg \to ZZH$ production. In Fig.~\ref{fig:dist-ZZH-pen-bx-tr}, 
we have shown the $p_T(H)$ distributions for penta, box, triangle, sum of their individual contributions, 
interference, and total at the 100 TeV collider  (FCC-hh). As can be seen, the box diagrams give the largest contribution, 
then comes the triangle contribution and the penta contributes the least. 
Like in $\gamma ZH$ case, the large box contribution is due to 
the light quarks in the loop.
Further, because of large {\it destructive} 
interference, the 
total contribution is smaller by about a factor of five than the box contribution. 
\\

\begin{figure}[H]
\begin{center}
\includegraphics[angle=0,width=0.48\linewidth]{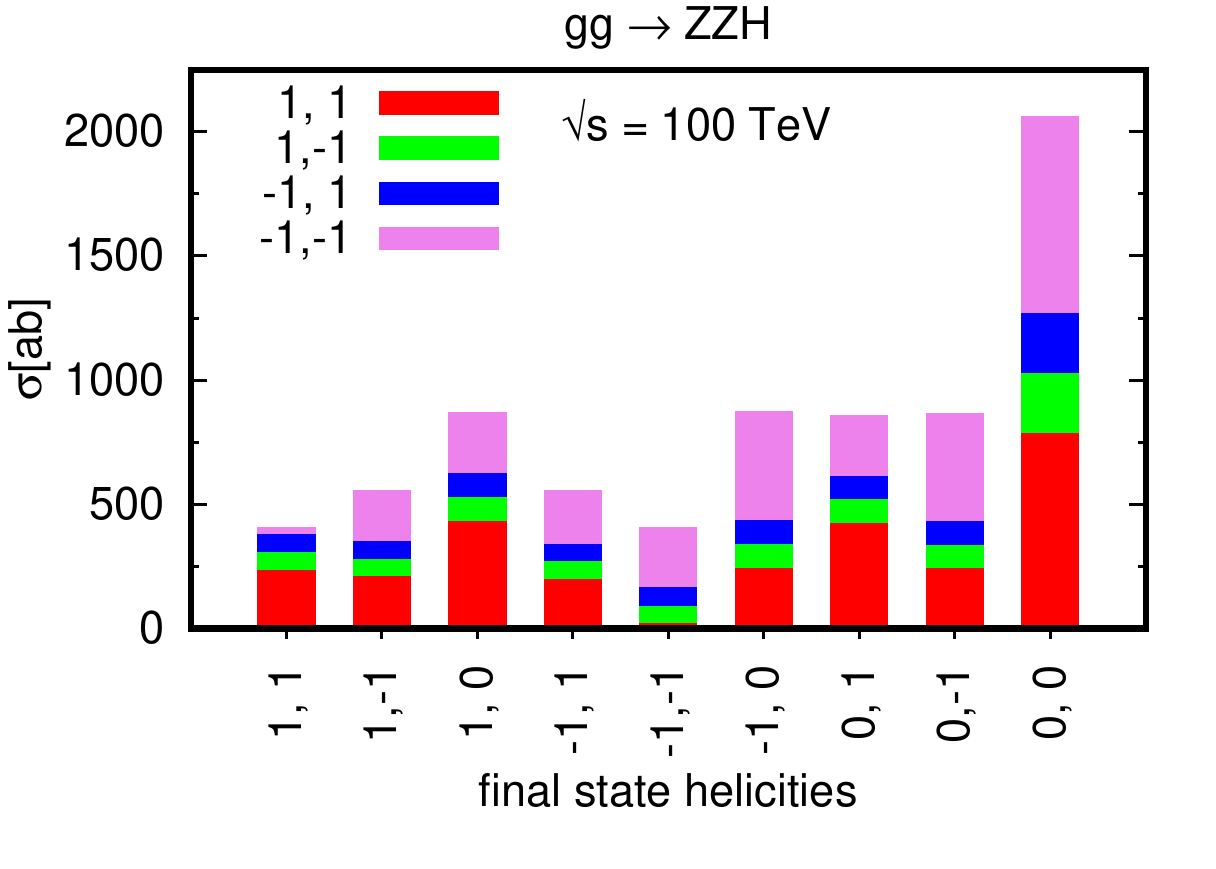}
\includegraphics[angle=0,width=0.48\linewidth]{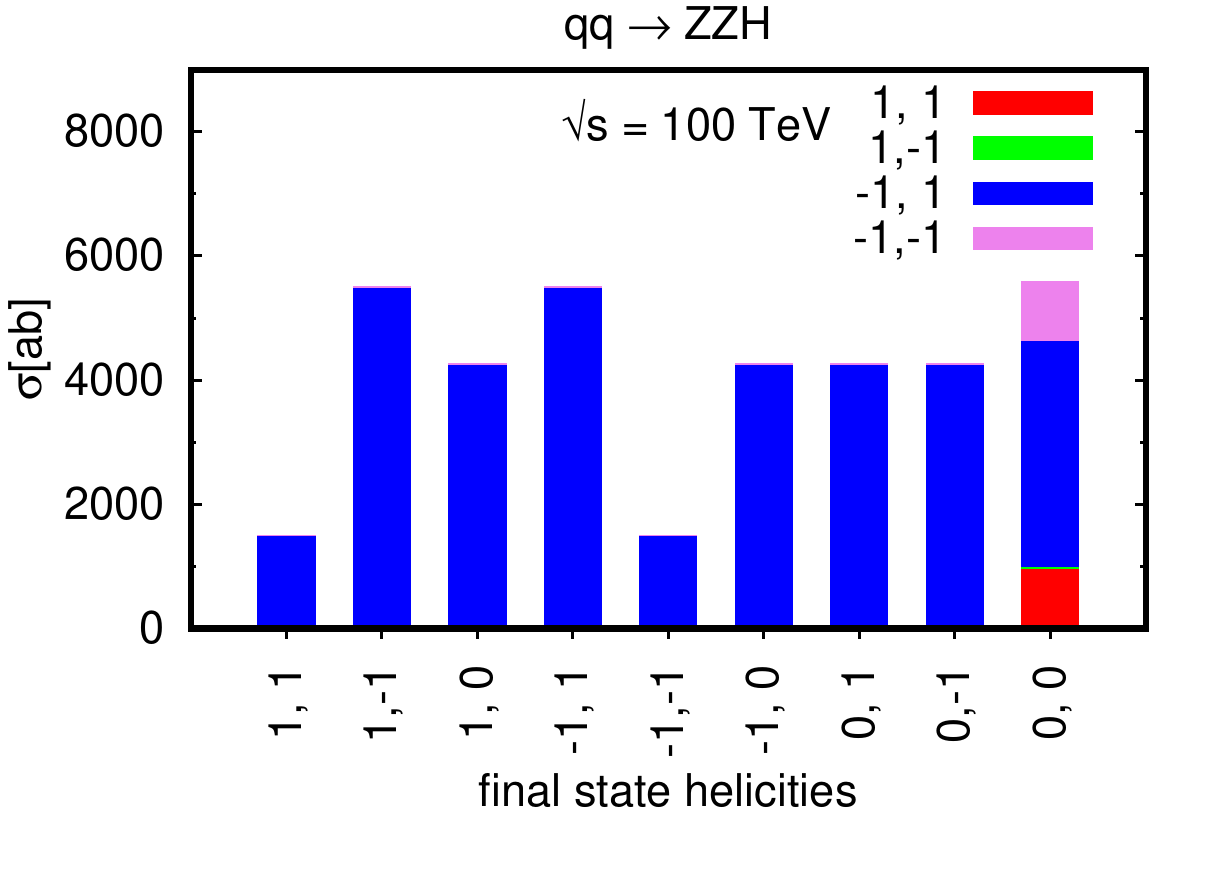}
\end{center}
\caption{LO cross section for $ZZH$ production in different helicity configurations in $gg$ (left) and $qq$ (right) channels. Legends correspond to different helicities of initial states.
}\label{fig:xs-hist-helicity-ZZH}
\end{figure}

 We have found that the top-quark contribution in $ggZZ^*$-type box diagram is quite significant despite the propagators suppression. This is due to the 
coupling of off-shell longitudinal $Z$ boson (effectively the Goldstone boson) with top-quark and it is  proportional to $m_t$~\footnote{The results for $ZZH$ process presented in the 
 conference proceeding~\cite{Shivaji:2016lnu} did not include top-quark contribution. We also fixed a bug in the code, numerical impact of which has 
 been found to be small.}. We show the effect of excluding the top-quark contribution in $ggZZ^*$-type box diagram (Fig.1(b)) 
 on $p_T(H)$ distribution in the right panel of Fig.~\ref{fig:dist-ZZH-pen-bx-tr}. As we expect, excluding top-quark contribution in $ggZZ^*$-type box diagram leads to non-unitary behavior in the full amplitude.
\\

In the left figure of Fig.~\ref{fig:dist-ZZH-QQ-GG}, we see that the shape of $p_T$ distribution for Higgs boson in the $gg$ and $qq$ channel processes is nearly same at 100 TeV collider (FCC-hh). The relative importance of the $gg$ channel over the $qq$ channel is visible in the tail.
In the right plot, we give $p_T(H)$ distribution 
combining $gg$ and $qq$ (NLO) contributions as the best prediction from our 
calculations. 
In the bottom panel of the plot, $R_2$ signifies the ratio of differential cross section from the $gg$ channel to that from NLO $qq$ channel process. The dashed line shows the ratio of corresponding total cross sections, which is 0.17.  At the tail of the distribution, we see the $gg$ channel contribution becomes further important, but there differential cross section itself is quite small.
\\


\begin{figure}[H]
\begin{center}
\includegraphics[angle=0,width=0.48\linewidth]{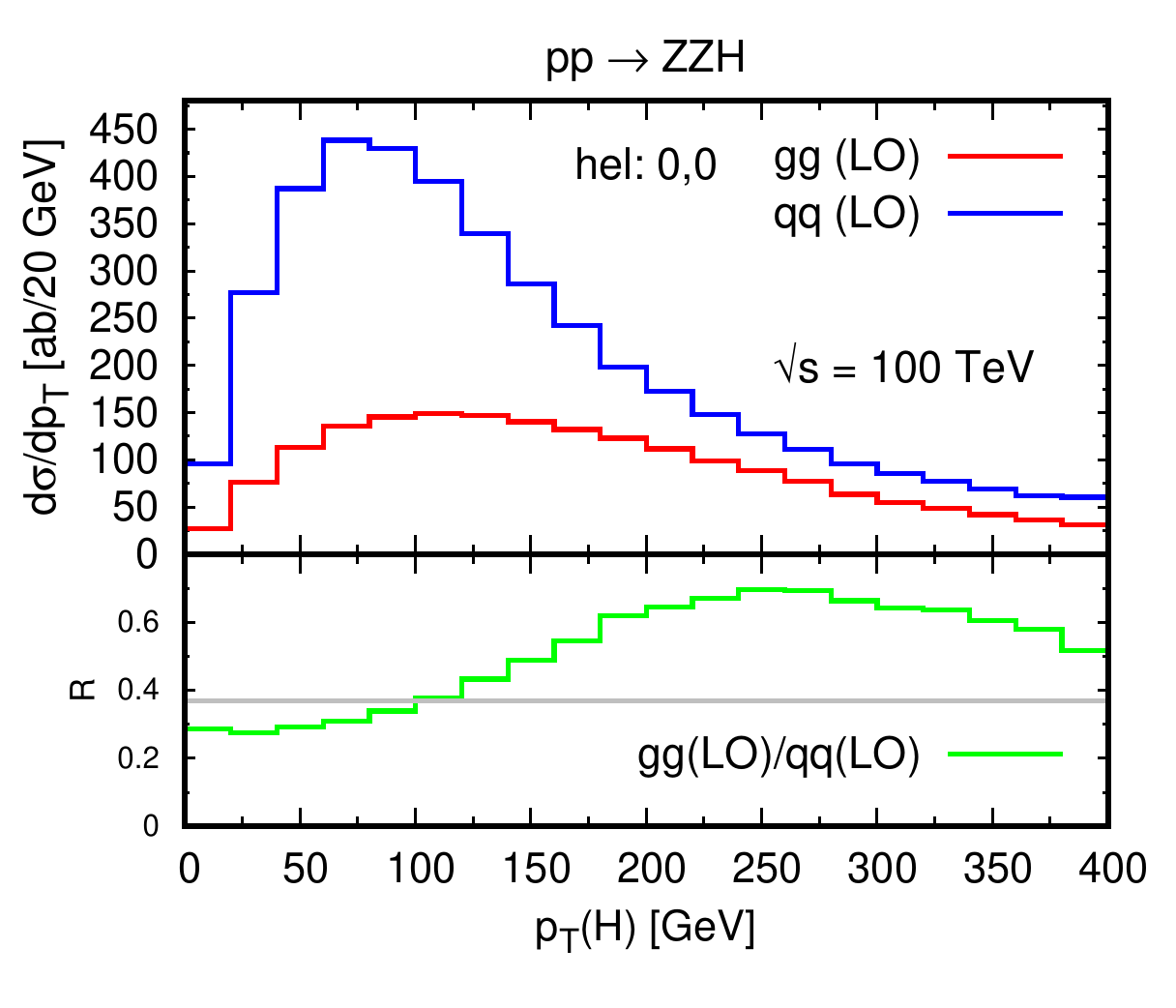}
\includegraphics[angle=0,width=0.48\linewidth]{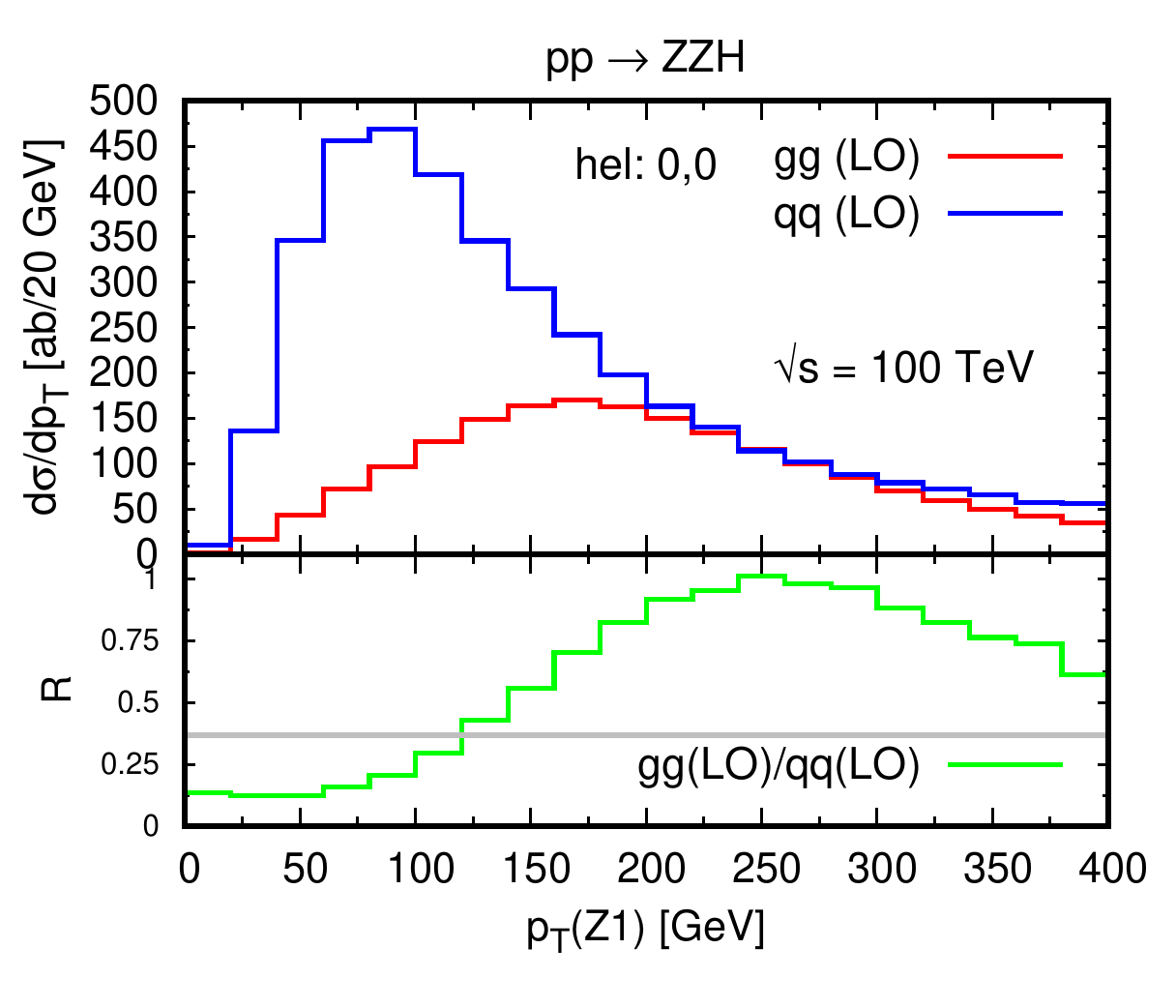}
\end{center}
\caption{The kinematic distributions from $gg$ and $qq$ channels when the final state $Z$-bosons are longitudinal. The ratio of the distributions from the two channels have been shown in the lower panel of each figure. In the right figure, $Z_1$ denotes the harder of two $Z$-bosons in $p_T$.}\label{fig:dist-helicity-ZZH}
\end{figure}

Once again we find that if we categorize events based on the helicity states 
of the two $Z$ bosons, the relative importance of the $gg$ channel contribution over 
the $qq$ channel contribution can be increased. 
From Fig.~\ref{fig:xs-hist-helicity-ZZH}, we see that in the $gg$ channel 
the longitudinal $Z$ bosons contribute the most, 
while in the $qq$ channel their transverse helicity states give dominant contribution. 
The relative cross section  of the $gg$ channel  with respect to the $qq$ channel is
about $20\%$. However, if we restrict ourselves to the case when both $Z$-bosons are 
longitudinally  polarized, then this ratio almost doubles. Since the cross section for these polarized states for the $gg$ channel is about 2000 ab, there will be enough events
to observe this process at a $100$ TeV machine.
At the distribution level, from the Fig. \ref{fig:dist-helicity-ZZH}, we observe that if we 
restrict ourselves to the contributions from 
the longitudinal $Z$ bosons with $p_T(H)$ beyond 150  GeV, the relative contribution of the $gg$ channel increases significantly. Experimentally, one may look at the signature $l^{+}l^{-}l^{+}l^{-}b\bar{b}$. This signature is obtained when  $Z \rightarrow  l^{+}l^{-} (l=e/\mu)$ and for $H \rightarrow \bar{b}b$. Taking into account the branching ratios, and
$b$-tagging efficiency, one may expect about 75 events at the FCC-hh collider (with 
$\rm{30\ ab}^{-1}$ integrated luminosity) from $gg$ channel and about
$210$ events from $qq$ channel. This is when both $Z$ bosons are longitudinally polarized. This number will go down when detection and kinematic-cut efficiency factors are included. However if in future,
one could use hadronic decay modes of a $Z$ boson to measure its
polarization, then the number of events would increase.
\\

As can be seen from Eq.~\ref{eq:ZZH}, the $gg$ channel depends on $\kappa_t,\ \kappa_V,\ \rm{and}\ \kappa_\lambda$.  We vary these $\kappa$'s by 10\% from their SM values.  The $gg$ channel strongly depends on both $\kappa_t\ \rm{and}\ \kappa_V$. In the $gg$ channel, $\pm$10\% change in $\kappa_t$ causes 68\% and -18\% variations in the cross section, respectively. And $\pm$10\% change in $\kappa_V$ causes 45\% and -28\% changes in the cross section, respectively. Similar variation in $\kappa_\lambda$ does not lead to much variation in the total cross section. Since this coupling is not yet well constrained, we will discuss it in detail in subsection~\ref{Remarks:HHH_and_HHVV}.

\begin{figure}[H]
\begin{center}
\includegraphics[angle=0,width=0.48\linewidth,height=0.35\linewidth]{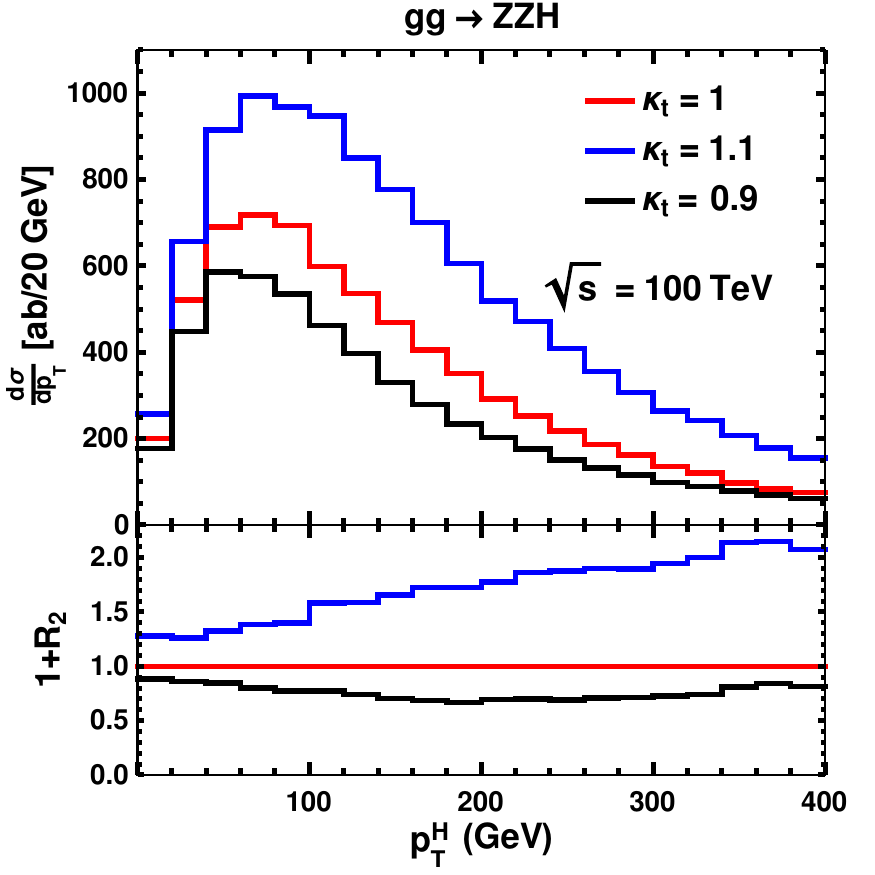}
\includegraphics[angle=0,width=0.48\linewidth,height=0.35\linewidth]{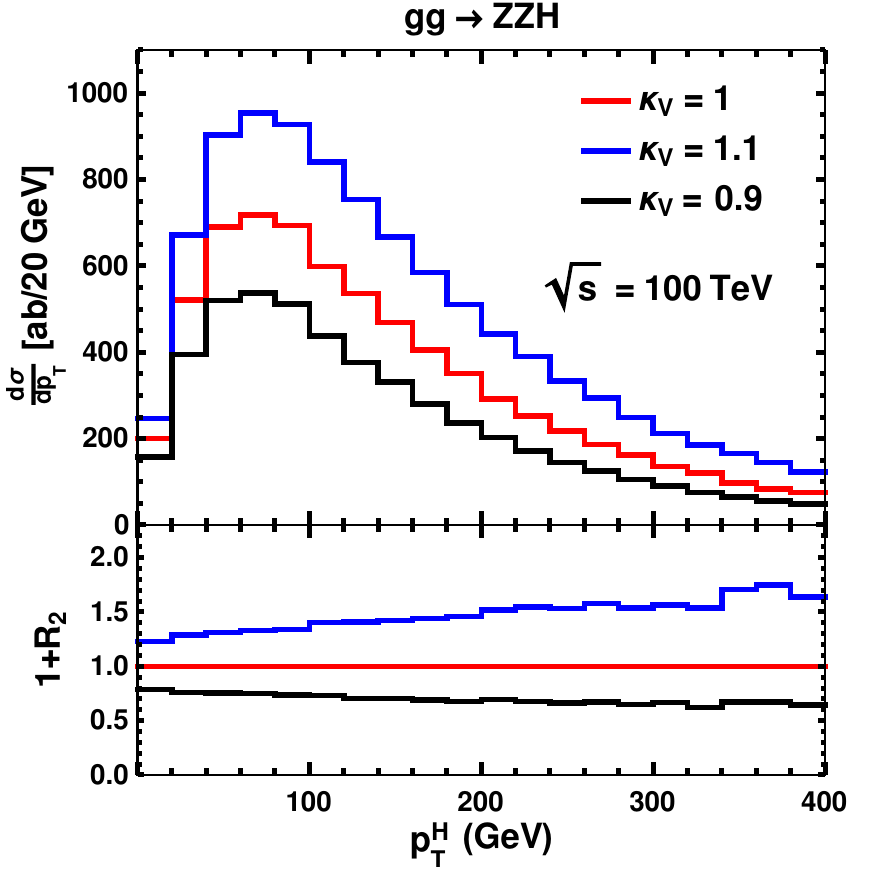}
\end{center}
\caption{Effect of anomalous values of $\kappa_{t}$ and $\kappa_{V}$ on 
$p_T(H)$ distribution for $ZZH$ production via the $gg$ channel. The 
lower panels display ratio of BSM and SM distributions.}  
 \label{fig:dist-ano-gg-ZZH}
\end{figure}

In Fig.~\ref{fig:dist-ano-gg-ZZH}, we display the effect of $\kappa_t$ and $\kappa_V$ on $p_T(H)$ distribution.  
We show the absolute distribution in the top panel, while in the bottom panel we show the ratio of distribution with anomalous coupling to that with the SM coupling. We can see that in the presence of anomalous $\kappa_t$ and $\kappa_V$, 
the shape of the distribution remains more or less same. 
However, due to non-trivial interference effects, the modifications in 
presence of anomalous couplings are not 
same in all the bins. 
We see that for $\kappa_t=1.1$ the cross section in the bins near 
tail of the distribution increases by a factor of 2. On the other 
hand for $\kappa_V=1.1$, the maximum change in the cross section 
is around 1.5. Thus tail of the distributions are 
more sensitive to modifications in couplings due to high scale new physics.
The $qq$ channel depends mainly on $\kappa_V$. However, as we have considered bottom quark contribution also, the $qq$ channel depends on $ \kappa_\lambda$ as well.  In the $qq$ channel, $\kappa_V$ comes as an overall factor both for LO and NLO amplitude, and so the effect of 10\% change in $\kappa_V$ causes around 20\% change in the cross section,
both at total and differential levels.  We find a very mild dependence on $\kappa_\lambda$.


\subsection{Predictions for $pp \to WWH$}\label{sec:results-WWH}

The cross section for this process is the largest among all the $VVH$ 
processes considered in this paper. In Tab.~\ref{table:xs-WWH}, we report 
the cross section predictions for $WWH$ process at different collider 
center-of-mass energies. 
The $gg$ channel contributions to $WWH$ production at 14, 27, and 100 TeV colliders are 290 ab, 1344 ab, and 17403 ab, respectively.
These numbers are roughly 2.3 times higher than $ZZH$ cross sections.
As regards scale uncertainties, the $gg \to WWH$ cross sections follow the same pattern as observed in $gg \to ZZH$. 
The corresponding values of the LO $qq$ channel cross sections are 8658, 23040, and 128000 ab, respectively\footnote{
Due to technical reasons in the NLO calculation  using {\tt MG5AMC\@NLO},
 the $qq$ results are provided in 4 flavor scheme.}. 
The ratio, $R_1$, is 
found to be 0.15, 0.19, and  0.43, respectively. Unlike $ZZH$ production, the contribution of the $gg$ channel is relatively smaller.

\begin{table}[H]
 \begin{center}
  \resizebox{0.8\columnwidth}{!}{%
  \begin{tabular}{|c|c|c|c||c||}
   \hline
   $\sqrt{\rm s}\;(\rm TeV)$ & $\sigma^{\tin{ \rm WWH,\,LO}}_{\tin{gg}}\;~[\rm ab]$ & $\sigma^{\tin{\rm WWH,\,LO}}_{\tin{qq}}\;~[\rm ab]$ & $\sigma^{\tin{\rm WWH,\,NLO}}_{\tin{qq}}\;[\rm ab]$ & ${R}_{1} $\\
   \hline
   & & & & \\
  14   & $290^{+27.6\%}_{-21.0\%}$ & $8658^{+0.3\%}_{-0.7\%}$ & $11220^{+1.5\%}_{-1.1\%}$ & 0.11  \\
   & & & & \\
  27   & $1344^{+22.5\%}_{-18.8\%}$ & $23040^{+2.1\%}_{-2.7\%}$ & $30090^{+1.7\%}_{-1.8\%}$  & 0.19 \\
   & & & & \\
  100   & $17403^{+20.6\%}_{-17.8\%}$ &   $128000^{+7.5\%}_{-8.1\%}$  & $167300^{+2.0\%}_{-3.3\%}$ &  0.44 \\
     & & & & \\     
   \hline
  \end{tabular}}
 \end{center}
	\caption{A comparison of different perturbative orders in QCD coupling contributing to $pp \to WWH$ hadronic cross section at $\sqrt{s}=$ 14, 27, and 100 TeV. The ratio $R_1$ defined in Eq.~\ref{eq:R1} quantifies the $gg$ channel contribution with respect to the $qq({\rm NLO})$ correction. The $qq$ results are reported in four flavor scheme.} \label{table:xs-WWH}
\end{table}

\begin{figure}[h]
\begin{center}
\includegraphics[angle=0,width=0.48\linewidth]{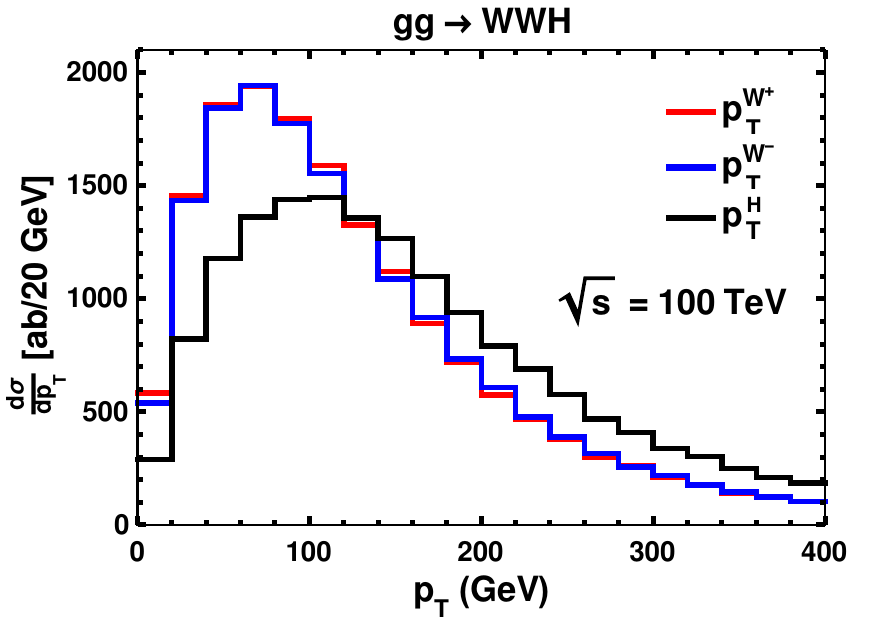}
\includegraphics[angle=0,width=0.48\linewidth]{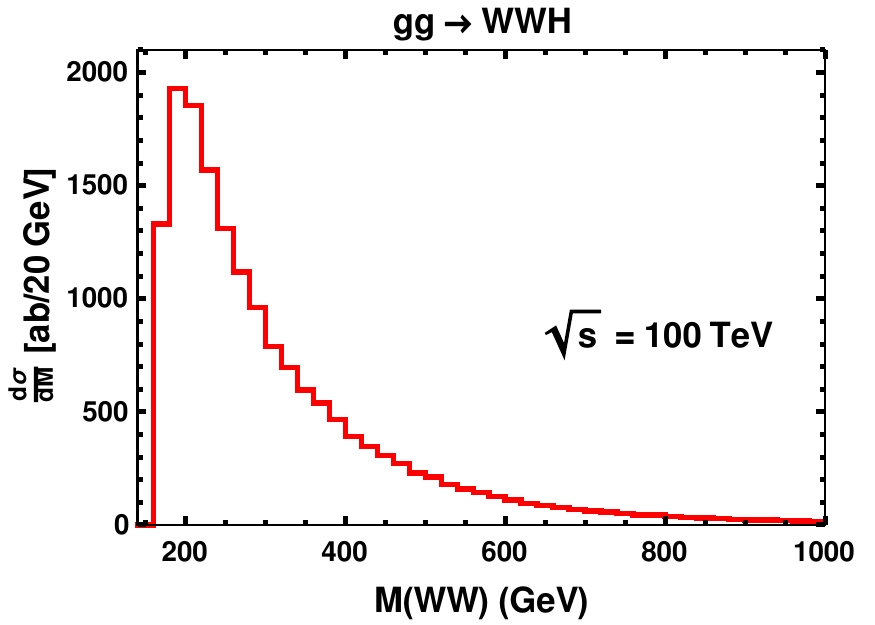}
\end{center}
\caption{ $p_T$ and $M(WW)$ distributions for $\GGWWH$ in the SM at the 100 TeV collider (FCC-hh).} 
\label{fig:dists-WWH-sm}
\end{figure}

In the left figure of Fig.~\ref{fig:dists-WWH-sm}, we can see that the $ p_T$ distribution of $W^{+}$ and $W^{-}$ overlap with each other, which is expected in the case of the $gg$ channel. The $p_T(H)$ distribution peaks around 100 GeV, and its fall in the tail is slower than that of $p_T(W^\pm)$ distributions. { In the right of  Fig.~\ref{fig:dists-WWH-sm}, the distribution for invariant mass of $W^+$ and $W^-$ has been shown, which peaks around 200 GeV.}

\begin{figure}[h]
\begin{center}
\includegraphics[angle=0,width=0.48\linewidth]{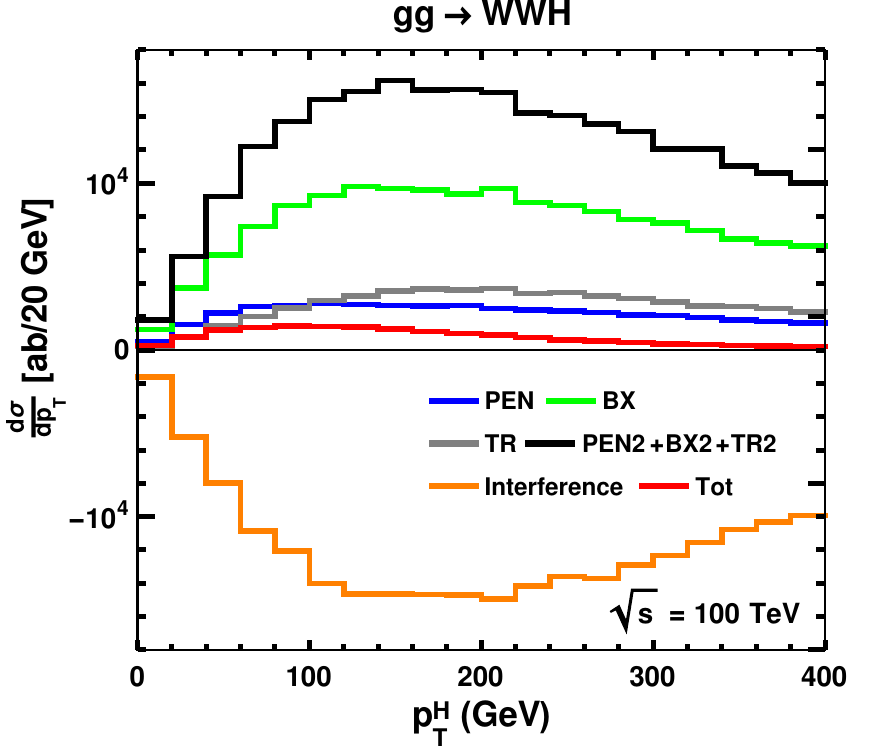}
\includegraphics[angle=0,width=0.48\linewidth]{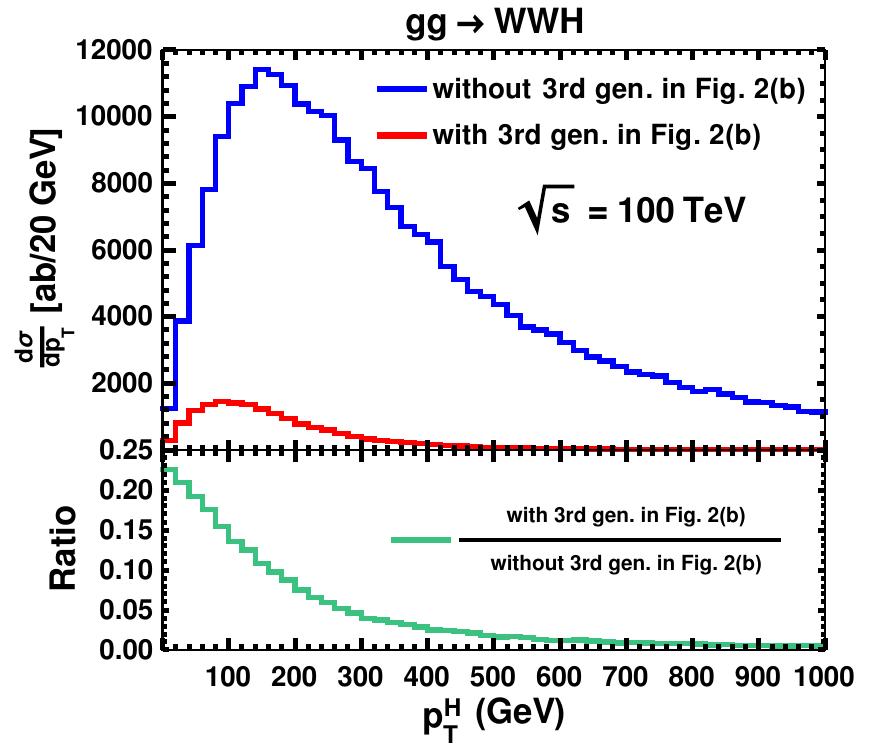}
\end{center}
\caption{Left : SM contribution of pentagon (blue), box(green), triangle (gray) diagrams, as well as their square sum, interference, and total contribution to $p_T(H)$ distributions in $\GGWWH$ at 100 TeV FCC-hh collider. 
Right: The effect of excluding third generation quark contribution from Fig. 2(b) to full amplitude.} \label{fig:dist-WWH-pen-bx-tr}
\end{figure}

Like $gg \to ZZH$ production case, in $gg \to WWH$ production also, interference of various diagrams plays a major role. On the left of Fig.~\ref{fig:dist-WWH-pen-bx-tr}, we have shown $ p_T(H)$ distributions 
for individual topologies as well as for their interference at a 100 TeV collider. The box contribution is the largest in all the bins while the 
pentagon contribution is the lowest beyond $p_T > 100$ GeV. The total contribution is much smaller than the box contribution because of strong destructive interference effect which is shown by orange line in the figure.

\begin{figure}[H]
\begin{center}
\includegraphics[angle=0,width=0.48\linewidth,height=0.35\linewidth]{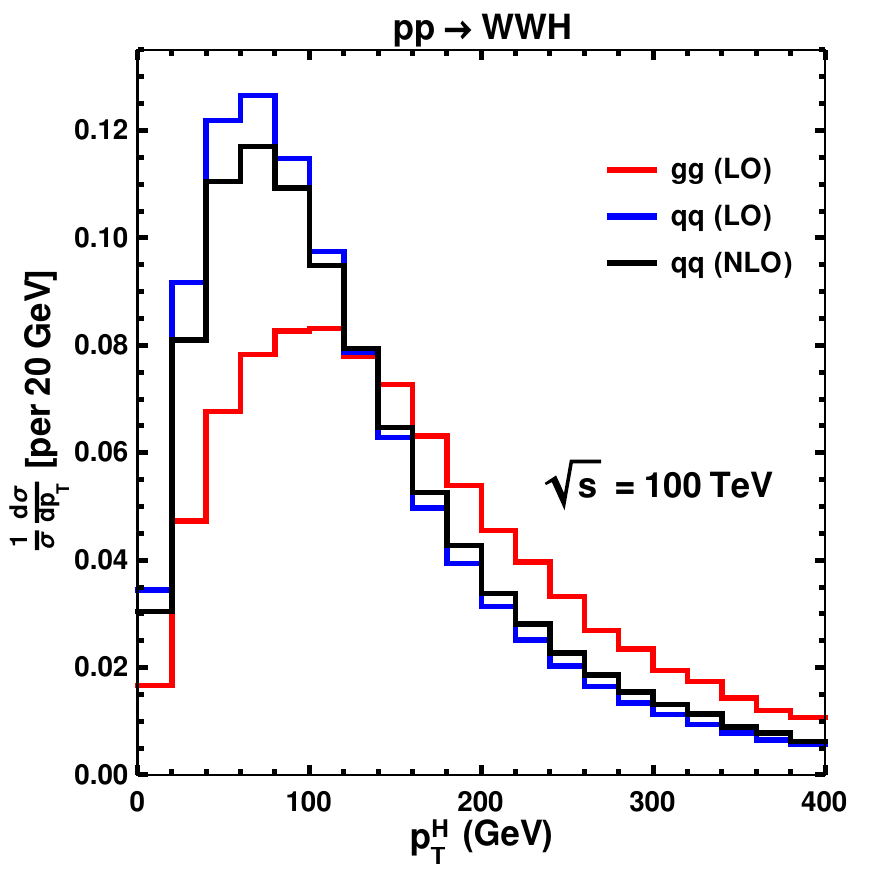}
\includegraphics[angle=0,width=0.48\linewidth,height=0.35\linewidth]{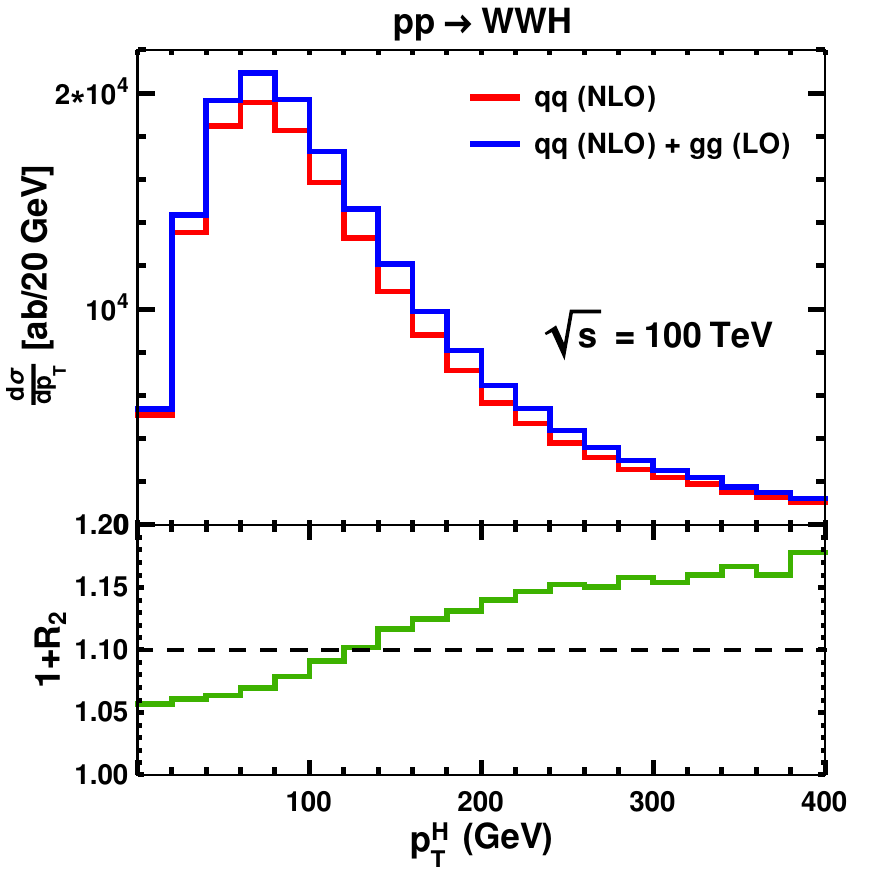}
\end{center}
	\caption{The left figure shows the normalized distribution for $p_T(H)$ in the $gg$ and $qq$ channel process. In the top panel of the right figure, we show the distribution due to $qq$ (NLO)+ $gg$ (LO) and $qq$ (NLO) production with $p_T(H)$. The lower panel shows their  ratio. Results do not include contribution of the $bb$ channel process.}  
 \label{fig:dist-WWH-QQ-GG}
\end{figure}

Due to the presence of top quark propagators in 
$ggWW^*$ type box diagram, one may naively 
think of a suppressed contribution from the 
third generation quarks at low $p_T(H)$. 
In Fig.~\ref{fig:dist-WWH-pen-bx-tr}, we show the effect of excluding the third 
generation quark contribution from the $ggWW^*$ type box diagram, on the $p_T(H)$ distribution.
Like in $gg \to ZZH$, the third generation quark contribution in $ggWW^*$ type box diagram is necessary for the unitarization of the full amplitude.
\\

In the left plot of Fig.~\ref{fig:dist-WWH-QQ-GG}, the normalized  $p_T$ distributions for Higgs boson in the $gg$ and $qq$ channel processes have been shown for 100 TeV collider (FCC-hh). The $p_T(H)$ distribution in 
the $gg$ channel peaks slightly on the harder side 
making the channel more relevant in higher $p_T(H)$ bins. 
To quantify it better we also plot the the ratio of distributions due to $qq$ (NLO) + $gg$ (LO) and $qq$ (NLO). At differential level 
the ratio varies between 1.05 and 1.18 compared to its value (1.1) for the total cross section. Once again, we find that the $gg$ channel contribution is more relevant at higher $p_T$ where its contribution reaches 18\%.


\begin{figure}[H]
\begin{center}
\includegraphics[angle=0,width=0.48\linewidth]{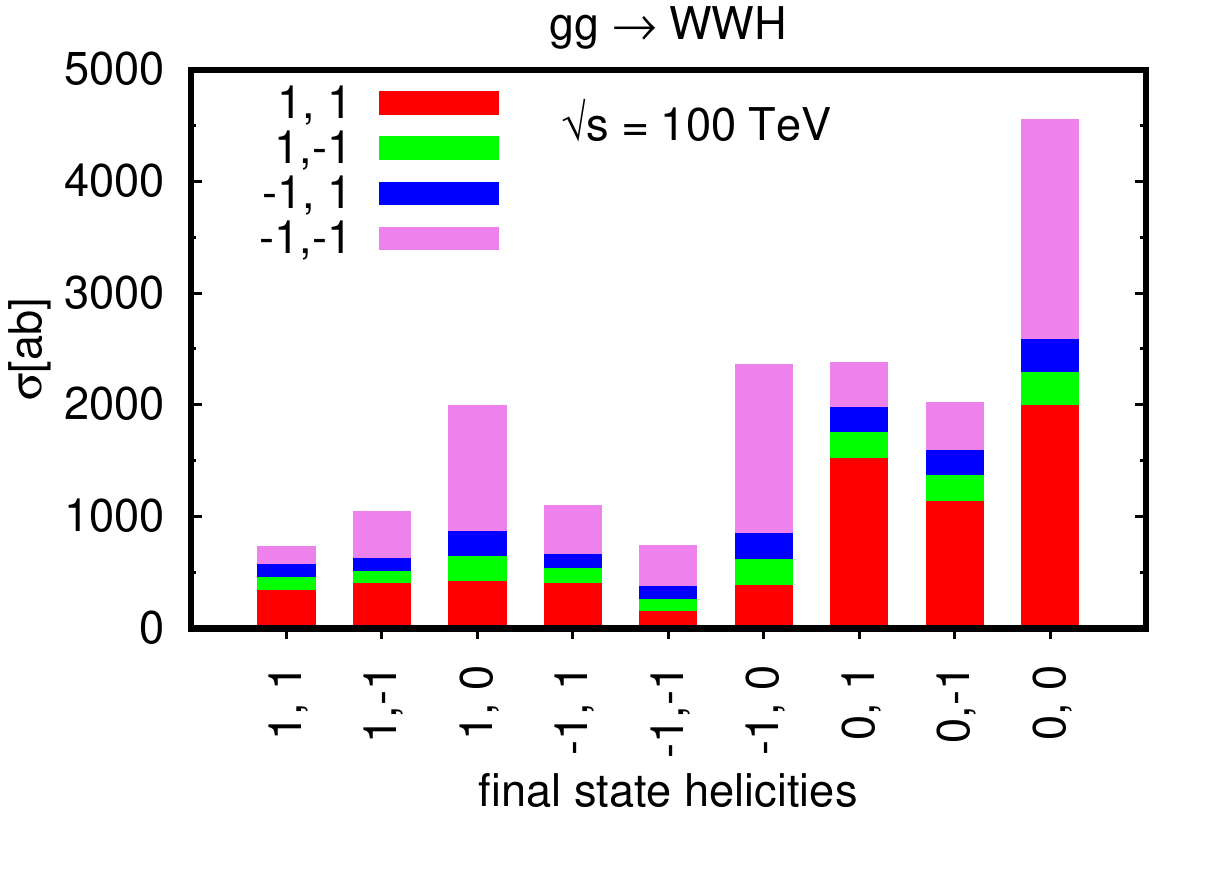}
\includegraphics[angle=0,width=0.48\linewidth]{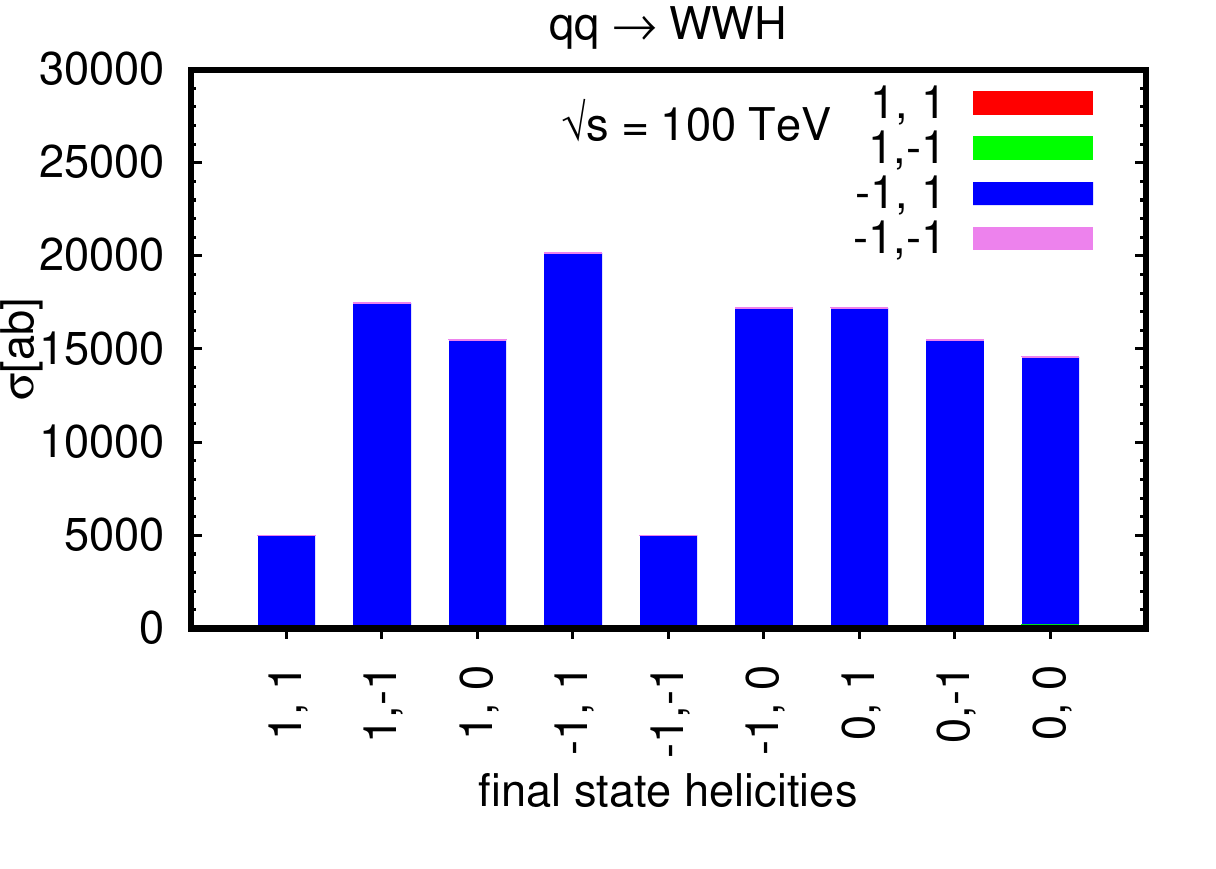}
\caption{LO cross section for $WWH$ production in different helicity configurations in $gg$ (left) and $qq$ (right) channels. Legends correspond to different helicities of initial states.}\label{fig:xs-hist-helicity-WWH}
\end{center}
\end{figure}

\begin{figure}[H]
\begin{center}
\includegraphics[angle=0,width=0.48\linewidth]{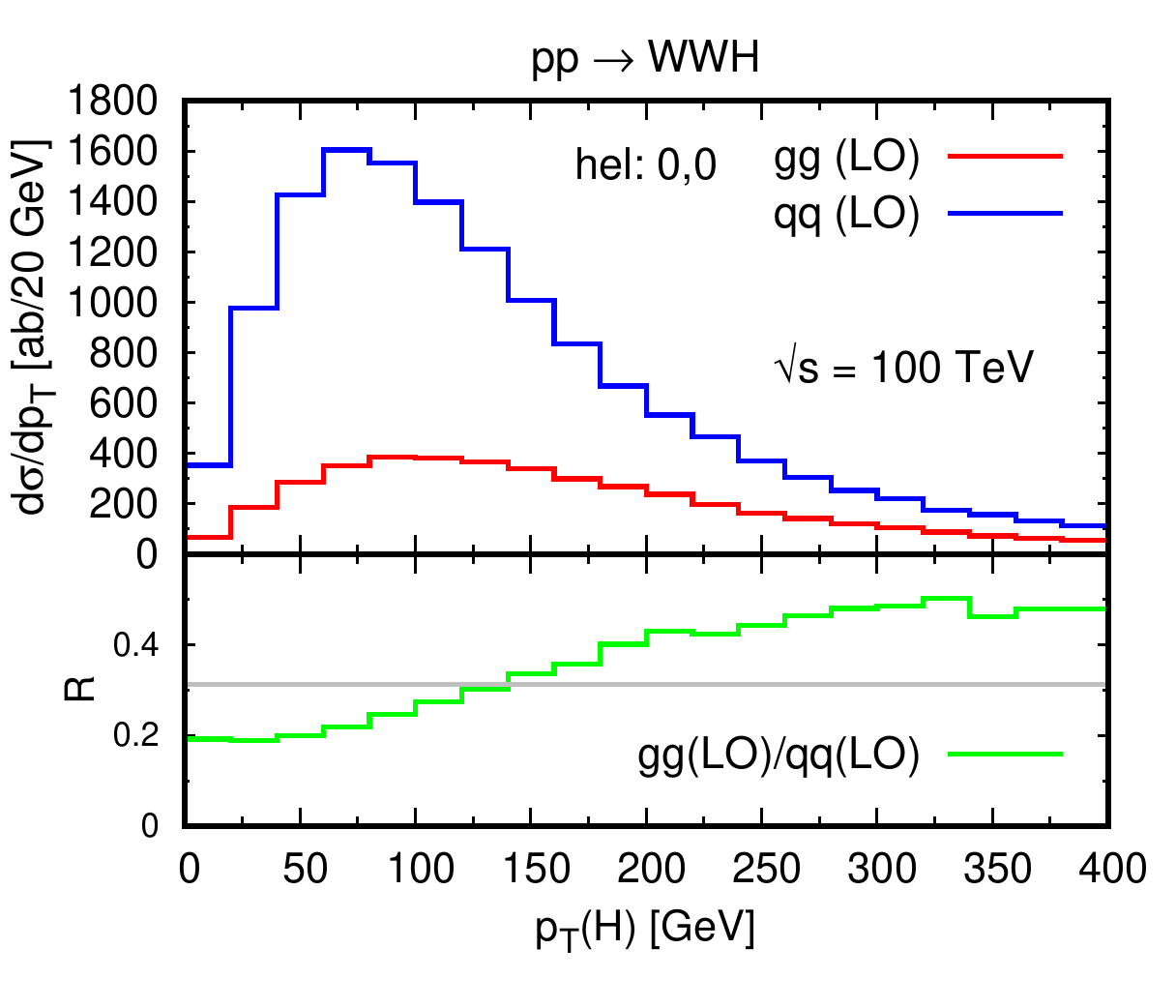}
\includegraphics[angle=0,width=0.48\linewidth]{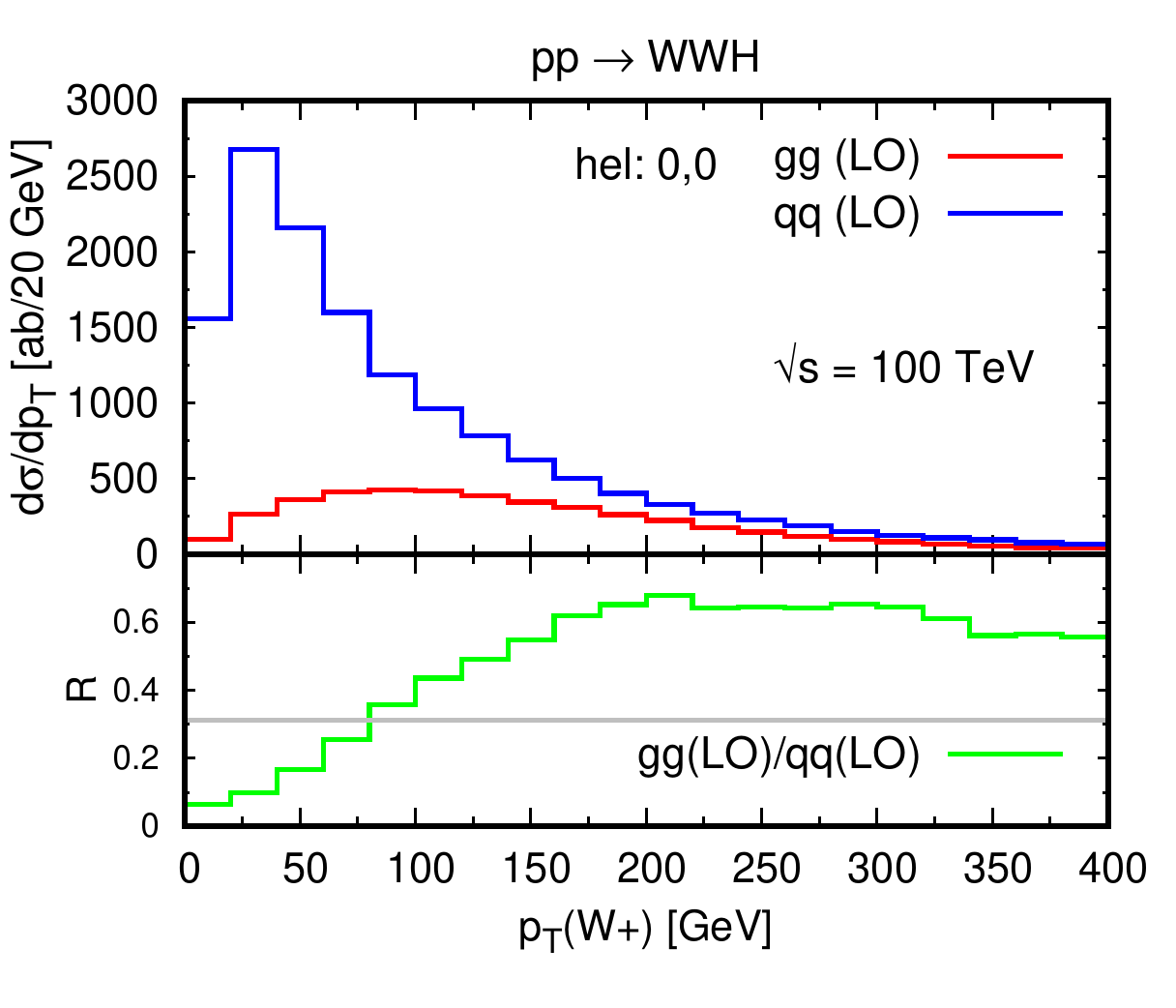}
\caption{Comparing the $gg$ and $qq$ channel contributions to $W^+W^-H$ for events with longitudinal $W$ bosons.}\label{fig:dist-helicity-WWH}
\end{center}
\end{figure}

Similar to the case of $ZZH$, for this process also, the cross section in the $gg$
channel is dominated by longitudinally polarized $W$-bosons (Fig.~\ref{fig:xs-hist-helicity-WWH}). The relative contribution of this channel is about $13\%$, with respect to the $qq$ channel. However,  when both $W$-bosons are longitudinally polarized, then this ratio increases to $32\%$. There will
also be enough events at a 100 TeV collider to observe the $gg$ channel contribution.
The relative 
contribution of the $gg$ channel over the $qq$ channel
can be further increased by requiring  the $p_T(W)$ to be beyond a certain value between 50 and 100 GeV,  see Fig. \ref{fig:dist-helicity-WWH}. Here also one may consider leptonic decay channel for $W$ bosons, as that will help in the measurement of its polarization. We consider $l^{+}\nu_{l}l^{-}{\bar{\nu}_l}b\bar{b}$ final state as the signature. Here, as
before $l = e/\mu$. In the literature, various techniques, including Neural Network methods have been discussed to measure the $W$ boson momentum \cite{Grossi:2020orx}. Taking into account the branching ratios and the $b$-tagging efficiency, one may expect about 1750 events from $gg$ channel and $5900$ events from the $qq$ channel at the FCC-hh collider with
$\rm{30\ ab}^{-1}$ integrated luminosity. The number of these events would change depending
on the detector and kinematic-cut efficiency factors.

\begin{figure}[H]
\begin{center}
\includegraphics[angle=0,width=0.45\linewidth,height=0.35\linewidth]{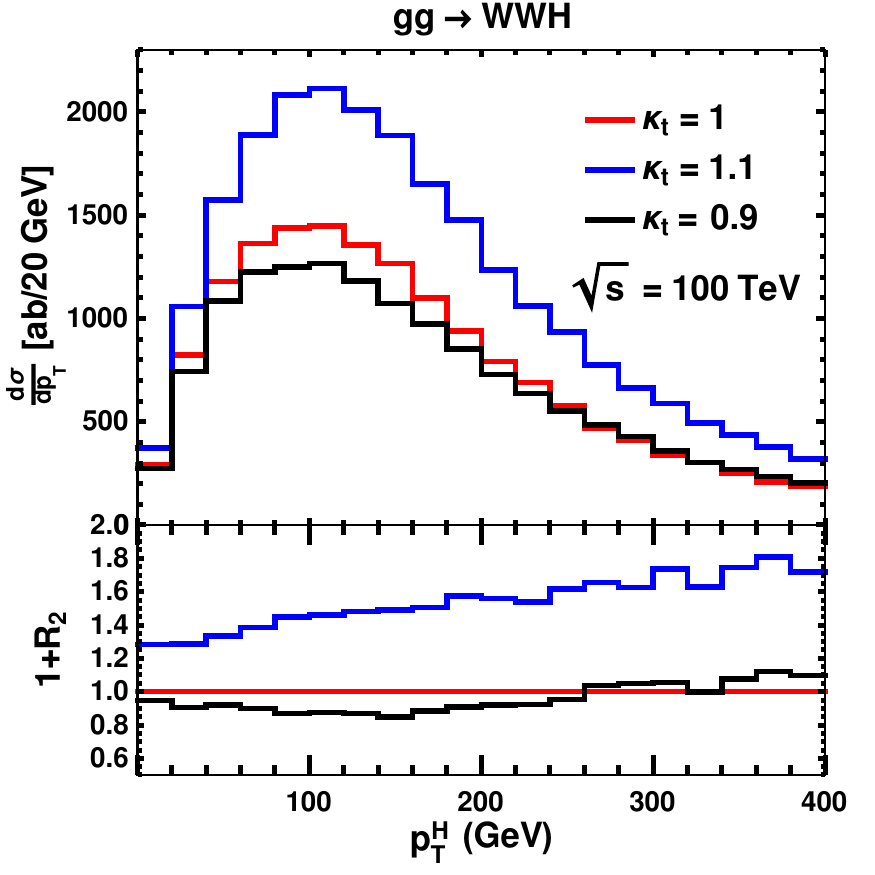}
\includegraphics[angle=0,width=0.45\linewidth,height=0.35\linewidth]{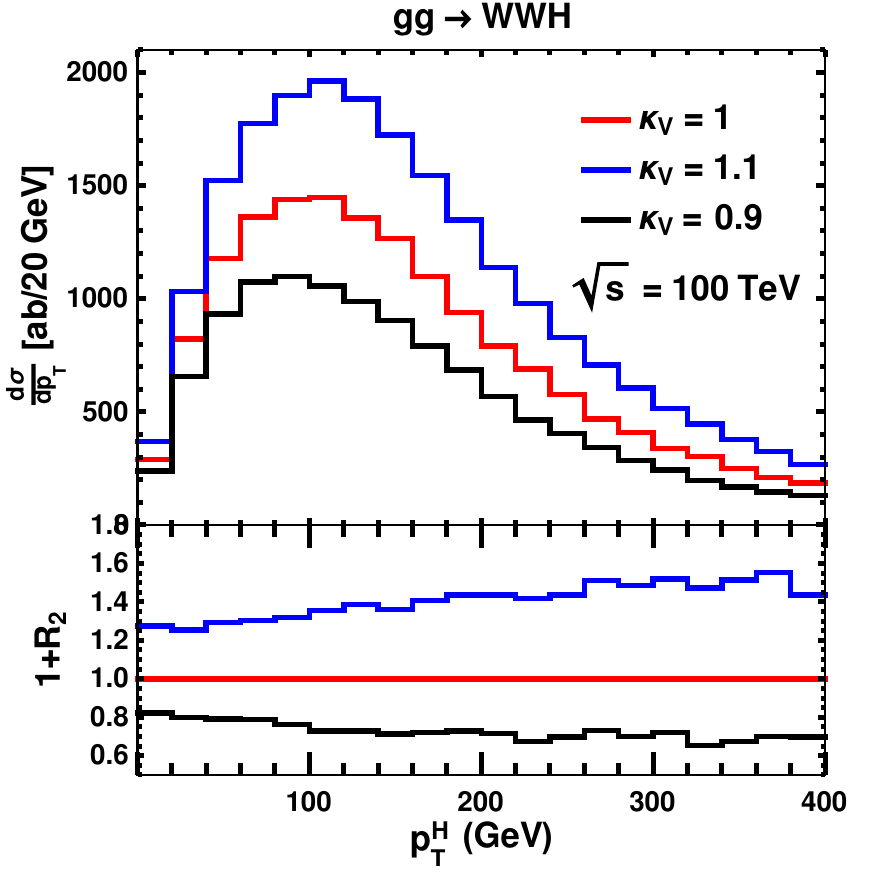}
\end{center}
\caption{Effect of anomalous values of $\kappa_{t}$ and $\kappa_{V}$ on $WWH$ production via the $gg$ channel. The upper panel shows absolute distribution, and the lower panel shows the ratio of the BSM and SM distributions.}  
 \label{fig:dist-ano-gg-WWH}
\end{figure}

Next, we  focus on the effect of anomalous couplings on the total and differential cross sections. The $gg$ channel depends on $\kappa_t,\ \kappa_\lambda,\ \rm{and}\ \kappa_V$ (see Eq.~\ref{eq:WWH}). 
We find that the channel is mostly sensitive to $\kappa_V$ 
and $\kappa_t$. For $\kappa_V=1.1(0.9)$ the cross section 
changes by about 38\%(-26\%). While, for $\kappa_t=1.1(0.9)$ the 
cross section changes by about 54\% (-3\%). The dependence on 
$\kappa_\lambda$ is found to be relatively small. 
In Fig.~\ref{fig:dist-ano-gg-WWH}, we show the effect of $\kappa_t$ and $\kappa_V$ on the $p_T(H)$ distribution for the $gg$ channel. We do not show the distribution for anomalous $\kappa_\lambda$ as its effect on cross section is very small for 10\% variation. We see that the shape remains more or less same in presence of anomalous couplings. 
We see that in the bins around 400 GeV, this ratio is around 1.5 for $\kappa_t=1.1$ and $\kappa_V=1.1$. For $\kappa_t=0.9$, the ratio remains close to 1 throughout all the bins and for $\kappa_V=0.9$, it is in the range 0.7--0.8. Similar to the case for $qq \to ZZH$, the $qq \to WWH$ cross section 
is also proportional to $\kappa_V^2$ at LO and NLO(QCD). So here as well, a 10\% change in $\kappa_V$ gives around 20\% obvious change in cross section, both at the total and differential levels.

\subsection{Remarks on anomalous $HHH$ and $HHVV$ couplings} \label{Remarks:HHH_and_HHVV}

{
We have seen that the gluon fusion $ZZH$ and $WWH$ processes 
are most relevant for BSM physics due to their large cross sections.  
We found that their cross sections do not 
change much for a 10\% variation in $\kappa_\lambda$. However, we know that this coupling is presently unconstrained 
by the experimental data. According to the future projections for 
HL-LHC, only values $\kappa_\lambda  \lesssim -2$ and $\kappa_\lambda \gtrsim 8$ can 
be ruled out~\cite{TheATLAScollaboration:2014scd}. In this range the cross section for $ZZH$ and 
$WWH$ processes in the $gg$ channel varies significantly. 
In fact, it can change maximum by a factor of 3.
This is shown in the left panel of Fig.~\ref{fig:dep-lam3-hhvv}. Notice that the $WWH$ process is more 
affected by anomalous $HHH$ coupling than $ZZH$ process. 
\\

\begin{figure}[h]
\begin{center}
\includegraphics[scale=0.6]{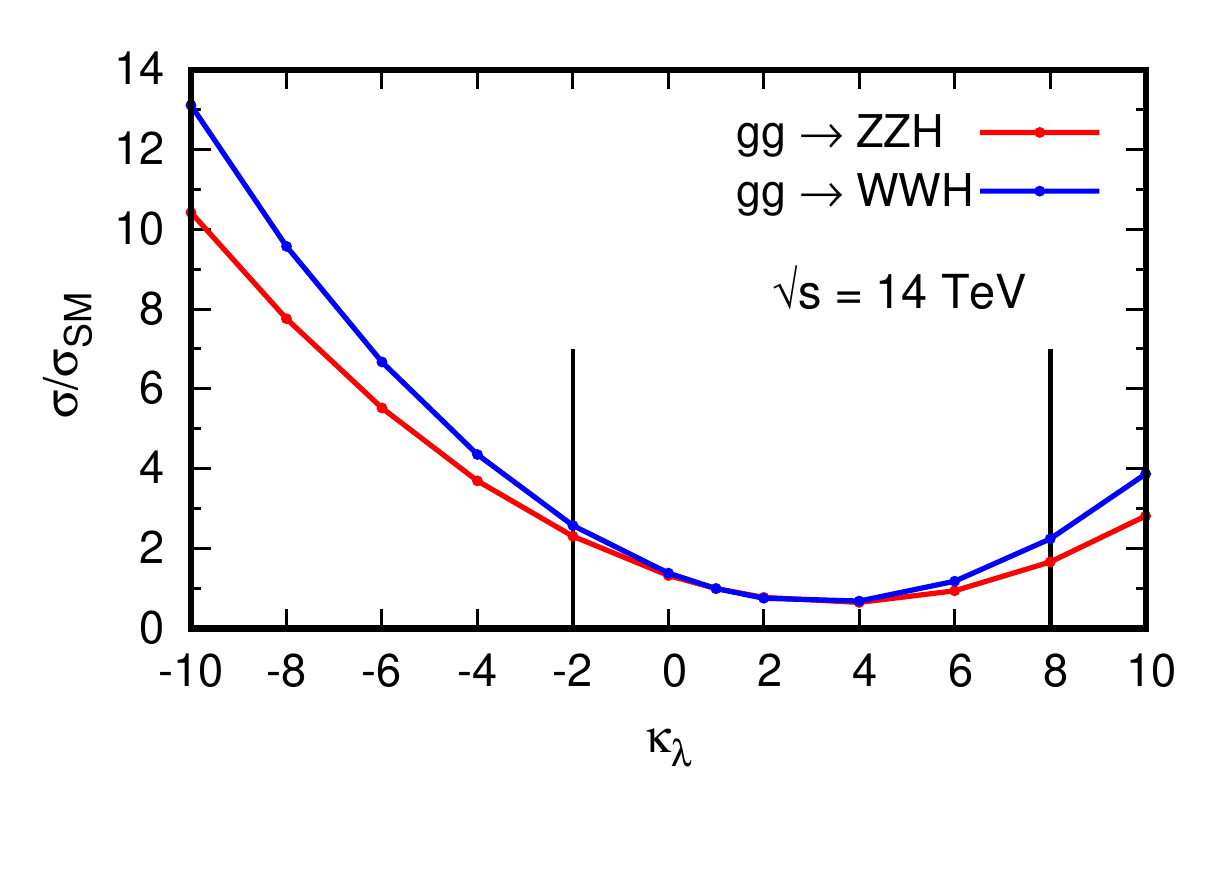}
\includegraphics[scale=0.6]{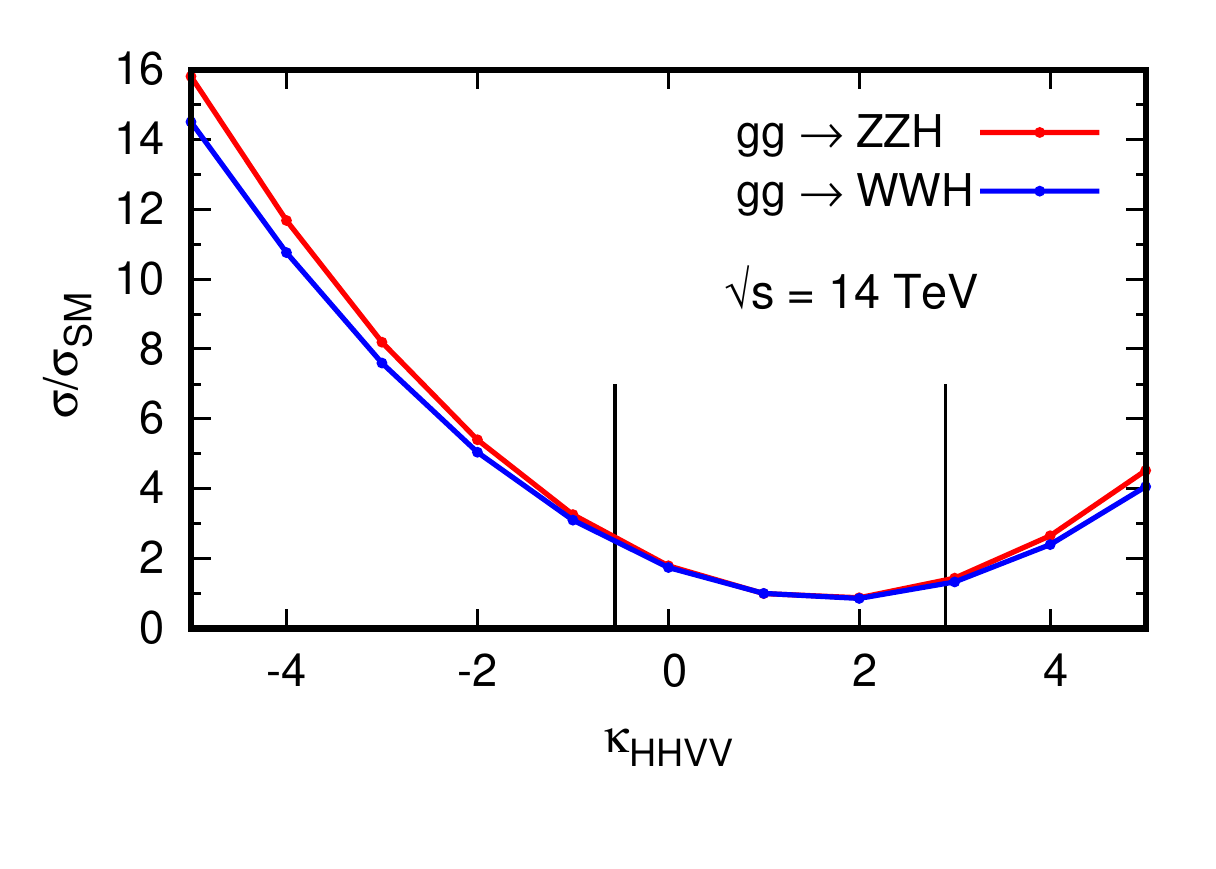}
\caption{Dependence of $gg \to ZZH, WWH$ cross sections on $HHH$ (left) and 
$HHVV$ (right) couplings at 14 TeV. The vertical lines in the left plot represent projected sensitivity on $\kappa_\lambda$ at HL-LHC and those on the right  represent current sensitivity on $\kappa_{HHVV}$ at the LHC.}
\label{fig:dep-lam3-hhvv}
\end{center}
\end{figure}

\begin{table}[h]
\begin{center}
  \resizebox{0.7\columnwidth}{!}{%
\begin{tabular}{|c|c|c|c||c|c|}
\hline 
Collider & $gg$ process & $c^{\kappa_\lambda}_1$ & $c^{\kappa_\lambda}_2$ & $c^{\kappa_{HHVV}}_1$ & $c^{\kappa_{HHVV}}_2$  \\
\hline
\hline
\multirow{2}{*}{14 TeV} & $ZZH$ & -0.275 & 0.053 & -0.458 & 0.335 \\
\cline{2-6}
& $WWH$ & -0.318 & 0.071 & -0.440 & 0.301\\
\hline
\hline
\multirow{2}{*}{100 TeV} & $ZZH$ & -0.256 & 0.046 & -0.563 & 0.772 \\
\cline{2-6}
& $WWH$ & -0.281 & 0.057 & -0.524 & 0.672\\
\hline
\end{tabular}}
\caption{$c^i_1$ and $c^i_2$ that appear in the definition of 
signal strengths for $gg \to ZZH, WWH$ processes at 14 TeV LHC and 100 TeV collider.}\label{table:c1-c2}
\end{center}
\end{table}

Although in SM model, the $HHVV (V = Z, W)$ coupling is correlated to the $HVV$ coupling, in presence of new physics this 
correlation may not exist. Keeping this possibility in mind, we have varied the $HHVV$ coupling independently\footnote{It should be noted that independent variation of $HVV$ and $HHVV$ couplings 
can be done systematically in an EFT framework which is beyond the scope of 
the present work.} and we find that the cross section changes 
very strongly. 
This is shown in the right panel of Fig.~\ref{fig:dep-lam3-hhvv}. We can see that the effect of the $HHVV$ coupling 
is relatively larger on $gg \to ZZH$ than on $gg \to WWH$. Close to SM values, the difference is negligible. According to a recent search 
for Higgs boson pair production via vector boson fusion carried out 
by the ATLAS collaboration using 126 $fb^{-1}$ data collected at 
13 TeV LHC, the allowed values of $\kappa_{HHVV}$ lie in the range 
(-0.56, 2.89) at 95\% confidence level~\cite{Aad:2020kub}.  
\\

The quantity plotted in Fig.~\ref{fig:dep-lam3-hhvv} is known as signal strength ($\mu$) which has been
utilized by experimentalists as observable for data analyses.
The signal strength for each process can be parametrized as  
\begin{eqnarray}
\mu = \frac{\sigma^{\rm BSM}}{\sigma^{\rm SM}} = 1 + c^i_1 (\kappa_i-1) + c^i_2(\kappa_i-1)^2,
\end{eqnarray}
where $\kappa_i = \kappa_\lambda, \kappa_{HHVV}$. In table~\ref{table:c1-c2}, we have 
provided the values of $c_1^i$ and $c_2^i$ for $ZZH$ and $WWH$ processes for the 14 TeV LHC and a 100 TeV pp collider. 
We note that $c_2^{k_\lambda}$ is smaller by an order of magnitude than $c_1^{k_\lambda}$, suggesting a strong interference effect mentioned before. Therefore, $c_2^{\kappa_\lambda}$ is relevant mostly for large values of $\kappa_i$. On the other hand, $c_2^{\kappa_{HHVV}}$ is of the same order as $c_1^{\kappa_{HHVV}}$. Since $c_1^i$ is negative, the cross section increase observed in the figures for $\kappa_i < 1$ is quite significant, which causes the (negative) lower bound on $\kappa$ to be tighter than the (positive) upper one. At a 100 TeV pp collider, while the other $c_i$s remain more or less same as that in 14 TeV collider, $c_2^{\kappa_{HHVV}}$ increases by around a factor of two, implying the possibility of a far more stringent bound on the $HHVV$ couplings. }
\\

\begin{figure}[H]
\begin{center}
\includegraphics[scale=0.6]{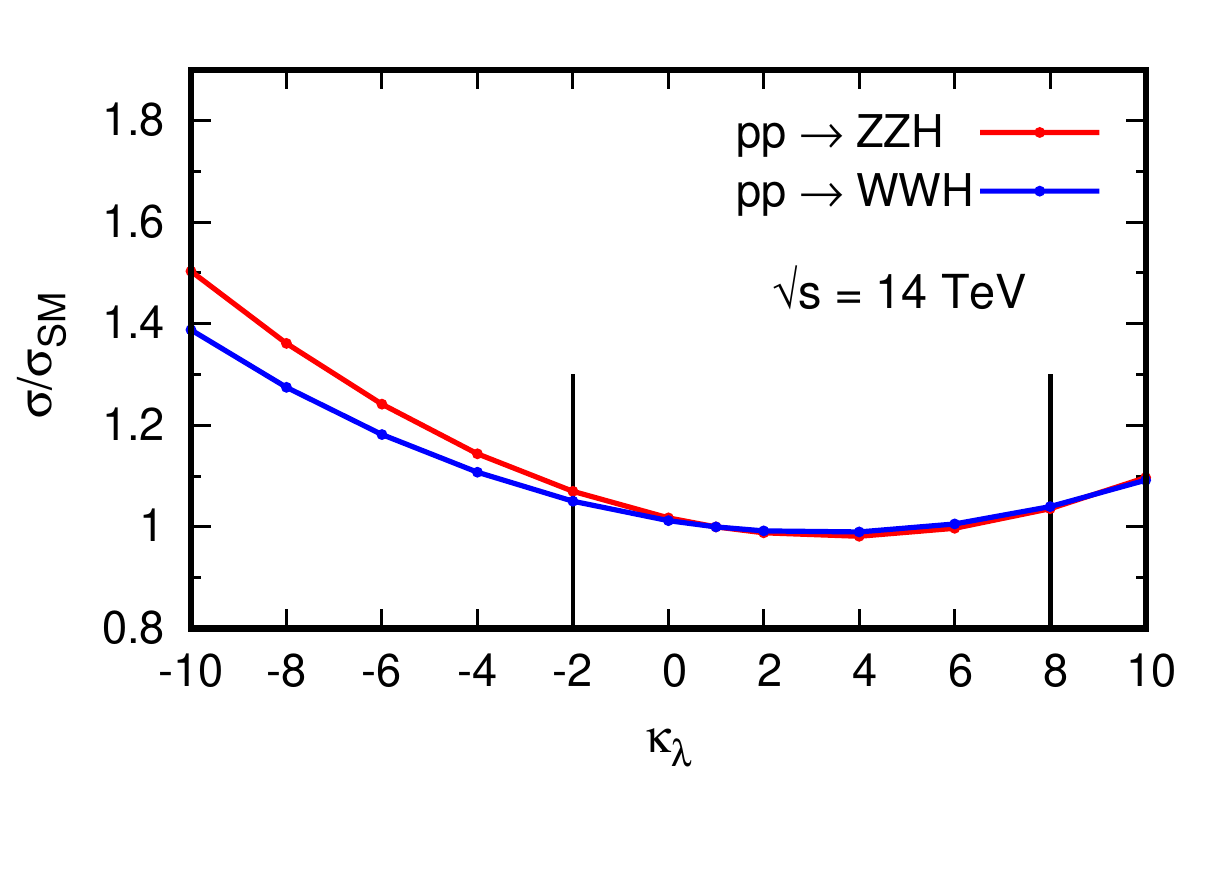}
\includegraphics[scale=0.6]{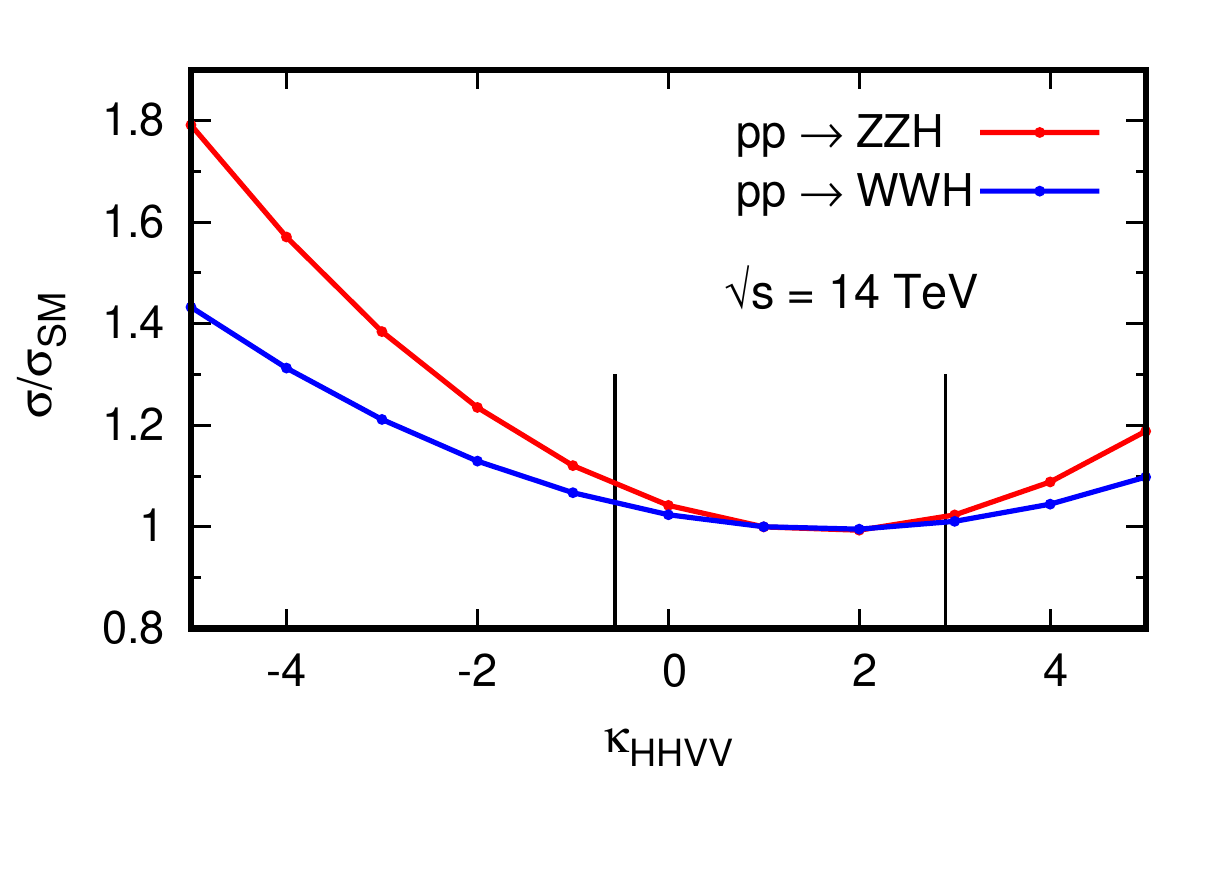}
\caption{Dependence of $pp \to ZZH, WWH$ cross sections on $HHH$ (left) and 
$HHVV$ (right) couplings at 14 TeV. The vertical lines in the left plot represent projected sensitivity on $\kappa_\lambda$ at HL-LHC and those on the right  represent current sensitivity on $\kappa_{HHVV}$ at the LHC.}
\label{fig:pp-dep-lam3-hhvv}
\end{center}
\end{figure}

Since the $gg$ fusion channel contribution to $ZZH$ and $WWH$ processes cannot be fully separated from the corresponding contributions from the $qq$ channel, the above result should be interpreted carefully. A realistic estimate of the BSM effects discussed above must include the contributions from $qq$ channel. Since $qq$ channel contributions are insensitive to $\kappa_\lambda$ and $\kappa_{HHVV}$, they can be seen as one of the major backgrounds to the gluon fusion processes. As we have pointed out, the measurement of the
polarization of the $W/Z$ boson can help in reducing this background. A systematic signal-background analysis is beyond the scope of the present work. For the benefit of the reader, in Fig~\ref{fig:pp-dep-lam3-hhvv}, we present the ratio $\sigma/\sigma_{\rm SM}$ for $pp \to ZZH, WWH$ which includes both $qq$ and $gg$ channel contributions as functions of $\kappa_\lambda$ and $\kappa_{HHVV}$. In obtaining these results only standard cuts mentioned in the previous sections have been applied. We can see that at the 14 TeV, the ratio of BSM and SM cross sections due to $qq+gg$ channels is significantly smaller than that due to $gg$ channel alone. Moreover, the $ZZH$ process turns out to be more affected by $\kappa_\lambda$ and $\kappa_{HHVV}$ than the $WWH$.
\\

To be more precise, we find that by changing $\kappa_\lambda$ in the range $(-2,8)$, the cross section for $ZZH$ process changes in the range $7-4\%$ at the 14 TeV. The corresponding change at the 100 TeV falls in the range of $20-8\%$. On the other hand, when changing $\kappa_{HHVV}$ in the range $(-0.56,2.89)$, the maximum cross section change in $ZZH$ process is found to be $\sim 8\%$ and $\sim~46\%$ at the 14 TeV and the 100 TeV, respectively. Again, we may mention that the polarization measurements would increase the
fraction of $gg$ channel events, thus increasing the dependence on $\kappa_{HHVV}$.

 \section{Conclusions}\label{sec:concl}

In this paper, we have considered production of $VV^\prime H$ ($\gamma\gamma H$, $\gamma ZH$, $ZZH$, and $WWH$) at 
 proton-proton colliders. 
We investigated the sensitivity  of these processes to various couplings of the Higgs boson, in particular to $HHH$ and $HHVV$ couplings 
which are practically unconstrained. 
Our main focus was the $gg$ channel contribution, which occurs at NNLO in $\alpha_s$.  The scale uncertainties on the total cross sections are found to be of the order of 20\%.
A number of checks like UV and IR finiteness and gauge invariance of the amplitudes with respect to the gluons 
have been performed to ensure the correctness of the calculation. 
At a 100 TeV collider, the cross sections for these processes via the $gg$ channel range from 0.2 fb to 17 fb, $gg \to WWH$ being the dominant channel among all. We have seen the $gg \to \gamma \gamma H$ and $gg \to \gamma Z H$ processes are insignificant background to $gg \to HH \to \gamma \gamma H$ and $gg \to HH \to \gamma Z H$, respectively.
\\

We have also compared the $gg$ channel contribution with the fixed order NLO QCD correction to $pp \to VV^\prime H$ in order to 
emphasize their relative importance.
For $\gamma \gamma H$ production, the $gg$ channel can be said to be the only production channel, as the $bb$ channel process contribution is negligibly small. 
At a 100 TeV collider, the $\GGAZH$ channel contribution is around 6\% of  the NLO QCD correction in the $qq$ channel. The $\gamma Z H$ production shows one interesting feature: with an increase in the $p_T$ cut on photon, the $qq$ channel contribution decreases faster than the $gg$ channel contribution.
At this collider, the contribution of the $gg$ channel to $ZZH$ production is as important as the fixed order QCD NLO correction to the $qq$ channel. On the other hand, the $\GGWWH$ channel cross section is around half the fixed order NLO QCD correction to the $qq$ channel. We have observed strong destructive interference effects among various classes of diagrams in $gg \to \gamma ZH, ZZH, WWH$. 
Besides total cross sections at the LHC, HE-LHC, and FCC-hh, we have obtained  relevant kinematic distributions at FCC-hh 
in the $gg$ channel. We find that the $p_T(H)$ spectrum from the $gg$ channel is harder than that from the $qq$  channel for ZZH and WWH productions.    
We have also shown that by selecting events based on the polarization of final state 
vector bosons, the relative contribution of the $gg$ channel over the $qq$  channel can be enhanced. 
\\

In addition to the SM results, effect of anomalous couplings ($\kappa_t$, $\kappa_V$, and $\kappa_\lambda$) for 
$Ht\bar{t}$, $HVV$, $HHVV$, and $HHH$ vertices have been studied in the kappa framework. 
We find that the new physics effects are quite important in $gg \to ZZH, WWH$ processes due to non-trivial interference effects in these processes. 
A 10\% change in $\kappa_t$ on the higher side can enhance the $gg \to ZZH$ and $WWH$ cross sections by 68\% and 54\%, respectively. Similar 
change in $\kappa_V$ enhances these cross sections by about 40\%. 
Unlike in $qq$ channels, the kinematic distributions in $gg$ 
channels display non-trivial changes in presence of new physics.  The dependence of the $g g$ channel on the $\kappa_V$ is stronger
than that of the $qq$ channel. By considering events with longitudinally polarized
vector bosons for the processes $ p p \to ZZH, WWH$, we can enhance the
fraction of the $gg$ channel contribution. This event sample will have even
stronger dependence on  $\kappa_V$.
Since the $HHH$ and $HHVV$ couplings are not well constrained, we have 
also considered larger independent variations  in $\kappa_\lambda$ and $\kappa_{HHVV}$. 
We find that the effect of $\kappa_{HHVV}$ on the cross section 
is much stronger than that of  $\kappa_\lambda$. Therefore the process 
$ p p \to ZZH, WWH$ with longitudinally polarized $Z$ and $W$ bosons
can help in determining the $HHVV$ coupling.

\clearpage 
\section*{Acknowledgements}
DS would like to acknowledge the use of HPC cluster facility, SAMKHYA, in Institute of Physics, Bhubaneswar. AS would 
like to acknowledge fruitful discussions with Xiaoran Zhao.


\appendix

\section{Comment on $Z$ mediated triangle diagrams in $gg \to WWH$}

\begin{figure}[H]
\centering	
\includegraphics[scale=0.6]{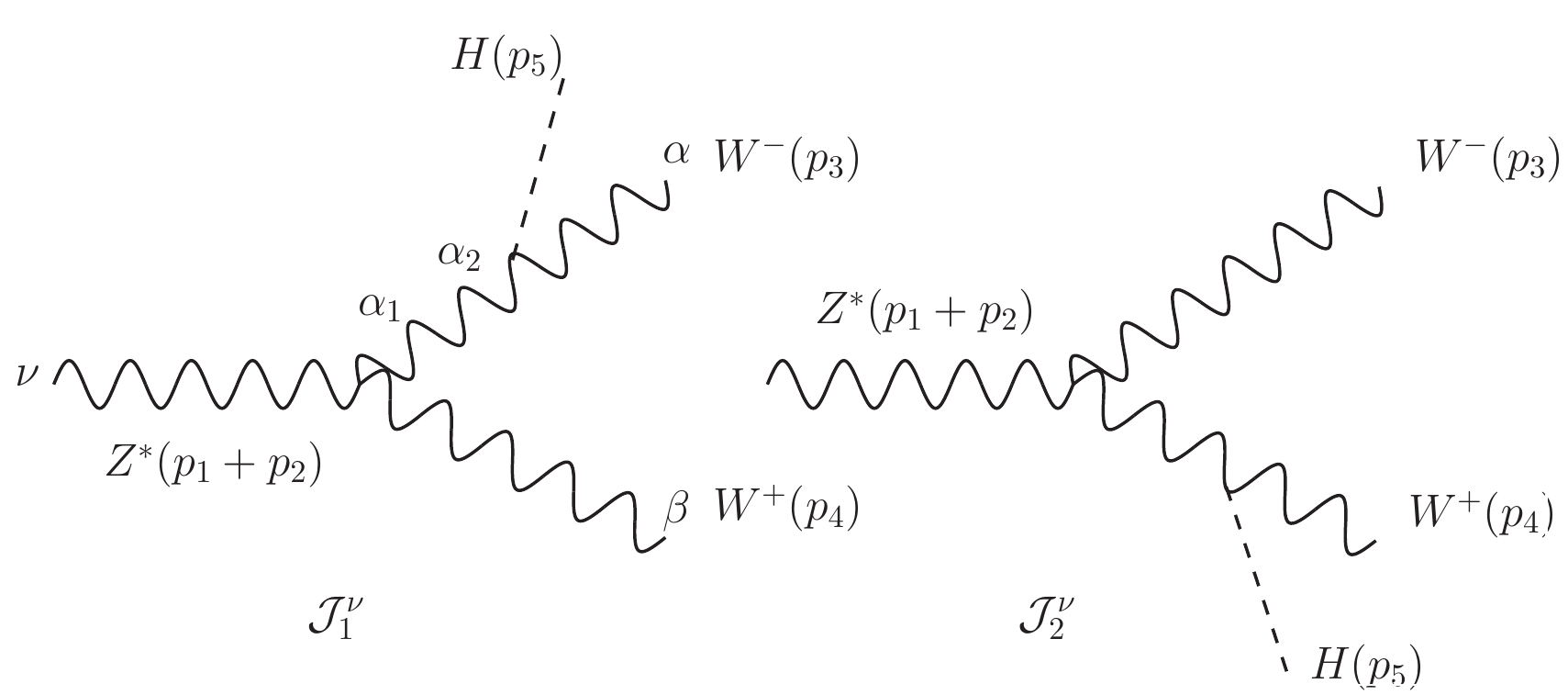}
	\caption{Currents attached to $Z^*$ in Fig.~\ref{fig:feyn-WWH-pen-bx-tr}(i). All the momenta are incoming. }\label{fig:ZHWW}
\end{figure}

It is a well known theorem due to Landau and Yang that a massive spin-1 particle 
cannot decay into two on-shell spin-1 massless particles~\cite{Landau:1948kw,Yang:1950rg}. The same theorem 
can be applied to argue that the $gg \to Z$ amplitude vanishes for on-shell $Z$ boson. 
This can be easily verified at LO using the on-shell conditions for the gluons and the $Z$ 
boson. In the past, we have shown that even if the $Z$ boson is off-shell, the LO 
$gg \to Z^*$ can vanish provided the off-shell $Z$ boson is linked to a conserved current~\cite{Shivaji:2013cca}. 
This is so because ${\cal M}^{\mu\nu\rho}(gg \to Z^*) \propto p_{Z^*}^\rho $. 
This result is useful for many $gg$ channel processes which receive contribution 
from such triangle topology. $gg \to WW$ is one such example~\cite{Binoth:2005ua,Campbell:2011cu}. 
In our case, $gg \to WWH$ is the process which depends on $Z$ mediated 
triangle diagrams. See Fig.~\ref{fig:feyn-WWH-pen-bx-tr} (h) and (i). We will explicitly show that Fig.~\ref{fig:feyn-WWH-pen-bx-tr}(i) does not 
contribute to the $gg \to WWH$ amplitude. For this we need to just prove that the sum of the 
currents shown in Fig.~\ref{fig:ZHWW} when contracted with the momentum $(p_1+p_2)^\nu$ vanishes. 

In the following derivation we use,
$p_1+p_2 = p_{12},~p_3+p_5=p_{35}$ and $p_4+p_5=p_{45}$. The polarization vectors of $W^-$ and $W^+$ are 
denoted by $e_3^\alpha(p_3)$ and $e_4^\beta(p_4)$, respectively.
We first calculate the contraction of current ${\cal J}_1$ with $p_{12}$. 

\begin{eqnarray}
	{\cal M}_1 = p_{12}^\nu {\cal J}_{1\nu} &=& p_{12}^\nu~ 
		     \Big( g_{\nu\alpha_1} (p_{12}-p_{35})_{\beta} + g_{\alpha_1\beta} (p_{35}-p_{4})_{\nu} 
			   + g_{\beta\nu} (p_{4}-p_{12})_{\alpha_1}
		     \Big) \times \nonumber \\ 
		    && \frac{-g^{\alpha_1\alpha_2}+p_{35}^{\alpha_1} p_{35}^{\alpha_2}/M_W^2}{p_{35}^2-M_W^2}~ g_{\alpha_2\alpha}~ 
		       e_3^{\alpha}(p_3)~ e_4^{\beta}(p_4) \\
		   &=& \Big( p_{12 \;\alpha_1}(p_{12}-p_{35}).e_4 + e_{4\;\alpha_1\;} p_{12}.(p_{35}-p_4) + 
		       p_{12}.e_4 (p_4-p_{12})_{\alpha_1}\Big) \times  \nonumber \\ 
		    && \frac{-e_3^{\alpha_1}+p_{35}^{\alpha_1} p_{35}.e_3/M_W^2}{p_{35}^2-M_W^2} \\ 
		   &=& \Big( -p_{12}.e_3 (p_{12}-p_{35}).e_4 - e_3.e_4 p_{12}.(p_{35}-p_4) 
		       - p_{12}.e_4 (p_4-p_{12}).e_3 \Big)/(p_{35}^2-M_W^2) \nonumber \\
		    && + \frac{p_{35}.e_3}{M_W^2(p_{35}^2-M_W^2)} \Big( p_{12}.p_{35} (p_{12}-p_{35}).e_4 
		       + p_{35}.e_4 p_{12}.(p_{35}-p_4) \nonumber \\ 
		    && - p_{12}.e_4 (p_4-p_{12}).p_{35} \Big)
\end{eqnarray}

Using momentum conservation $p_{12} = -p_{35}-p_4$ and transversality conditions $e_3.p_3 = e_4.p_4 = 0$, we get
\begin{eqnarray}
	{\cal M}_1 &=& \Big( -2 p_{35}.e_4 p_{45}.e_3 + e_3.e_4 (p_{35}^2-p_4^2) 
	               + p_{35}.e_4 (p_4+p_{45}).e_3\Big)/(p_{35}^2-M_W^2) \nonumber \\ 
		    && + \frac{p_{35}.e_3}{M_W^2(p_{35}^2-M_W^2)} \Big( 2 p_{35}.e_4(p_{35}+p_4).p_{35} 
		       - p_{35}.e_4 (p_{35}^2-p_4^2) - p_{35}.e_4 (2p_4+p_{35}).p_{35} \Big) \nonumber \\ \\
		    &=& \frac{-p_{35}.e_3 p_{35}.e_4 + e_3.e_4 (p_{35}^2-p_4^2)}{(p_{35}^2-M_W^2)} + 
		        \frac{p_{35}.e_3}{M_W^2(p_{35}^2-M_W^2)} p_{35}.e_4 p_4^2
\end{eqnarray}
Using on-shell condition $p_4^2=M_W^2$, we arrive at 
\begin{eqnarray}
	{\cal M}_1 &=& e_3.e_4\label{eq:M1} . 
\end{eqnarray}
Following similar steps, it can be shown that contraction of current ${\cal J}_2$ with $p_{12}$ leads to, 
\begin{eqnarray}
	{\cal M}_2 = p_{12}^\nu {\cal J}_{2\nu} &=& -e_3.e_4\label{eq:M2}. 
\end{eqnarray}
Combining equations \ref{eq:M1} and \ref{eq:M2} we obtain the desired result: ${\cal M}_1 + {\cal M}_2 =0$.
Thus we have proved that indeed the current associated with $Z^*$ in Fig.~\ref{fig:ZHWW}  is a conserved current and 
therefore the triangle amplitude for Fig.~\ref{fig:feyn-WWH-pen-bx-tr}(i) vanishes for each quark flavor in the loop. It can be 
verified explicitly that the current associated with $Z^*$ in Fig.~\ref{fig:feyn-WWH-pen-bx-tr}(h) is not a conserved current and 
therefore it does give non-vanishing contribution to $gg \to WWH$ amplitude.

\newpage


\input{\jobname.bbl}

\clearpage

\end{document}

%% file: VVH-draft_final.bbl
\providecommand{\href}[2]{#2}\begingroup\raggedright\endgroup